\pdfoutput=1
\documentclass[manuscript]{acmart}
\setcopyright{none}
\renewcommand\footnotetextcopyrightpermission[1]{}
\PassOptionsToPackage{table,xcdraw}{xcolor}
\usepackage{pifont}
\usepackage{colortbl}
\usepackage{booktabs, multirow, tabularx}

\definecolor{lightgreen}{RGB}{204,255,204}
\definecolor{lightred}{RGB}{255,204,204}
\definecolor{lightgrey}{RGB}{224,224,224}
\usepackage{float}
\usepackage{subcaption}
\usepackage{tabularx}
\usepackage{enumitem}
\usepackage{multirow}
\usepackage{makecell}
\usepackage{tikz}
\usetikzlibrary{arrows.meta}
\newlength{\luU}\newlength{\luA}\newlength{\luB}\newlength{\luC}\newlength{\luD}\newlength{\luE}
\newlength{\luBC}\newlength{\luH}\newlength{\luXb}\newlength{\luXc}\newlength{\luXd}\newlength{\luXe}
\definecolor{luaccent}{HTML}{4C5BD4}   
\definecolor{luweak}{HTML}{EAECFB}     
\definecolor{luframe}{HTML}{DFE3EE}    
\definecolor{luwarn}{HTML}{D55E00}     
\definecolor{luwarm}{HTML}{FBF1E6}     
\definecolor{lurow}{HTML}{F1F2F4}      
\definecolor{lumark}{HTML}{5A6472}
\definecolor{luink}{HTML}{2F3AA3}      
\definecolor{lurule}{HTML}{D5D9E0}     
\definecolor{lugrey}{HTML}{8A94A3}     
\definecolor{luweakly}{HTML}{62A8D2}   
\tikzset{lupanel/.style={draw=luframe,line width=0.4pt,rounded corners=2pt,fill=white,inner sep=4pt,anchor=north west,align=left,font=\scriptsize,execute at begin node={\hyphenpenalty=10000\exhyphenpenalty=10000}},
         lubadge/.style={circle,fill=luaccent,text=white,font=\sffamily\bfseries\tiny,inner sep=0pt,minimum size=7.5pt,anchor=south west},
         lucard/.style={rounded corners=2pt,inner sep=2.5pt,align=left,font=\scriptsize,text width=#1,execute at begin node={\hyphenpenalty=10000\baselineskip=8pt}},
         lucard/.default=60pt,
         luexplicit/.style={draw=luaccent,line width=0.8pt,fill=luweak},
         luderived/.style={draw=lugrey,line width=0.4pt,dashed,dash pattern=on 1.6pt off 1.1pt,fill=white}}
\newcommand{\lubadge}[2]{\node[lubadge] at ([yshift=2pt]#1.north west) {#2};}
\newcommand{\lutype}[1]{{\tiny\textcolor{lugrey}{\textsc{#1}}}}
\newcommand{\luweak}{{\tiny\textcolor{luweakly}{$\bullet$ weakly satisfied}}}
\newcommand{\lumarker}{\textcolor{lumark}{\scriptsize$\blacktriangleright$\,(A)--(D)}}
\usepackage{graphicx}
\usepackage{subcaption}

\definecolor{TellingColor}{RGB}{94,197,186}
\definecolor{SellingColor}{RGB}{233,170,114}
\definecolor{ParticipatingColor}{RGB}{217,124,97}
\definecolor{DelegatingColor}{RGB}{149,163,201}
\definecolor{BaselineColor}{RGB}{163,163,163}
\definecolor{applegrey}{RGB}{99, 99, 102}

\newcommand{\card}[1]{\emph{#1}}
\newcommand{\lvl}[1]{\textsc{#1}}
\usepackage{xcolor}
\definecolor{patAuthoring}{HTML}{FCE4D6}
\definecolor{patMixed}{HTML}{EDEDE4}
\definecolor{patGatekeeping}{HTML}{D8EDE2}
\newcommand{\patbox}[2]{{\setlength{\fboxsep}{1pt}\colorbox{#1}{\hspace{2pt}\rule[-0.5ex]{0pt}{1.9ex}#2\hspace{2pt}}}}
\newcommand{\patA}[1]{\patbox{patAuthoring}{#1}}
\newcommand{\patM}[1]{\patbox{patMixed}{#1}}
\newcommand{\patG}[1]{\patbox{patGatekeeping}{#1}}

\begin{document}

\title{IntentWeaver: Sustaining Coherence across Evolving Intents in Human–AI Co-Writing}

\author{Yuanhao Zhang}
\email{yzhangiy@connect.ust.hk}
\affiliation{%
  \institution{The Hong Kong University of Science and Technology}
  \city{Hong Kong}
  \country{China}}

\author{Xinru Cheng}
\email{chengxinru0313@gmail.com}
\affiliation{%
  \institution{Peking University}
  \city{Beijing}
  \country{China}}

\author{Baiqiao Zhang}
\email{bzhangcw@connect.ust.hk}
\affiliation{%
  \institution{The Hong Kong University of Science and Technology}
  \city{Hong Kong}
  \country{China}}

\author{Xinqi Zhang}
\email{xzhang22@scu.edu}
\affiliation{%
  \institution{Santa Clara University}
  \city{Santa Clara}
  \state{California}
  \country{United States}}

\author{Assem Zhunis}
\email{azhunis@connect.ust.hk}
\affiliation{%
  \institution{The Hong Kong University of Science and Technology}
  \city{Hong Kong}
  \country{China}}

\author{Shuai Ma}
\email{mashuai@iscas.ac.cn}
\affiliation{%
  \institution{Institute of Software, Chinese Academy of Sciences}
  \city{Beijing}
  \country{China}}

\author{Xiaojuan Ma}
\email{mxj@cse.ust.hk}
\affiliation{%
  \institution{The Hong Kong University of Science and Technology}
  \city{Hong Kong}
  \country{China}}

\renewcommand{\shortauthors}{Zhang et al.}

\begin{abstract}
  Writers hold intents at many levels, from overarching goals to concrete wording, and evolve them by periodically popping back up from the text to higher-level intents. Generative AI co-writing disrupts this practice. Dense drafts capture attention and trap writers in local editing; higher-level intents stall while lower-level ones drift from them unnoticed, a breakdown we term intent incoherence. Our formative study traced this breakdown from attention capture through silent drift to failed reconciliation. We present IntentWeaver, which supports intent evolution while sustaining coherence in human-AI co-writing. It derives an intent hierarchy from the draft and writer interactions, evaluates each intent's fulfillment and cross-level coherence at every change, and propagates consequences across intent levels. In a within-subjects study (N=24) against a baseline, participants using IntentWeaver revisited and evolved higher-level intents more often, with fewer unnoticed deviations, and produced texts better realizing them. We further discuss design implications for emergent intents.
\end{abstract}

\ccsdesc[500]{Human-centered computing~Interactive systems and tools}
\ccsdesc[300]{Human-centered computing~Empirical studies in HCI}

\keywords{AI-assisted Writing, Intent Communication, Human-AI Collaboration}

\begin{teaserfigure}
  \centering
  \includegraphics[width=\textwidth]{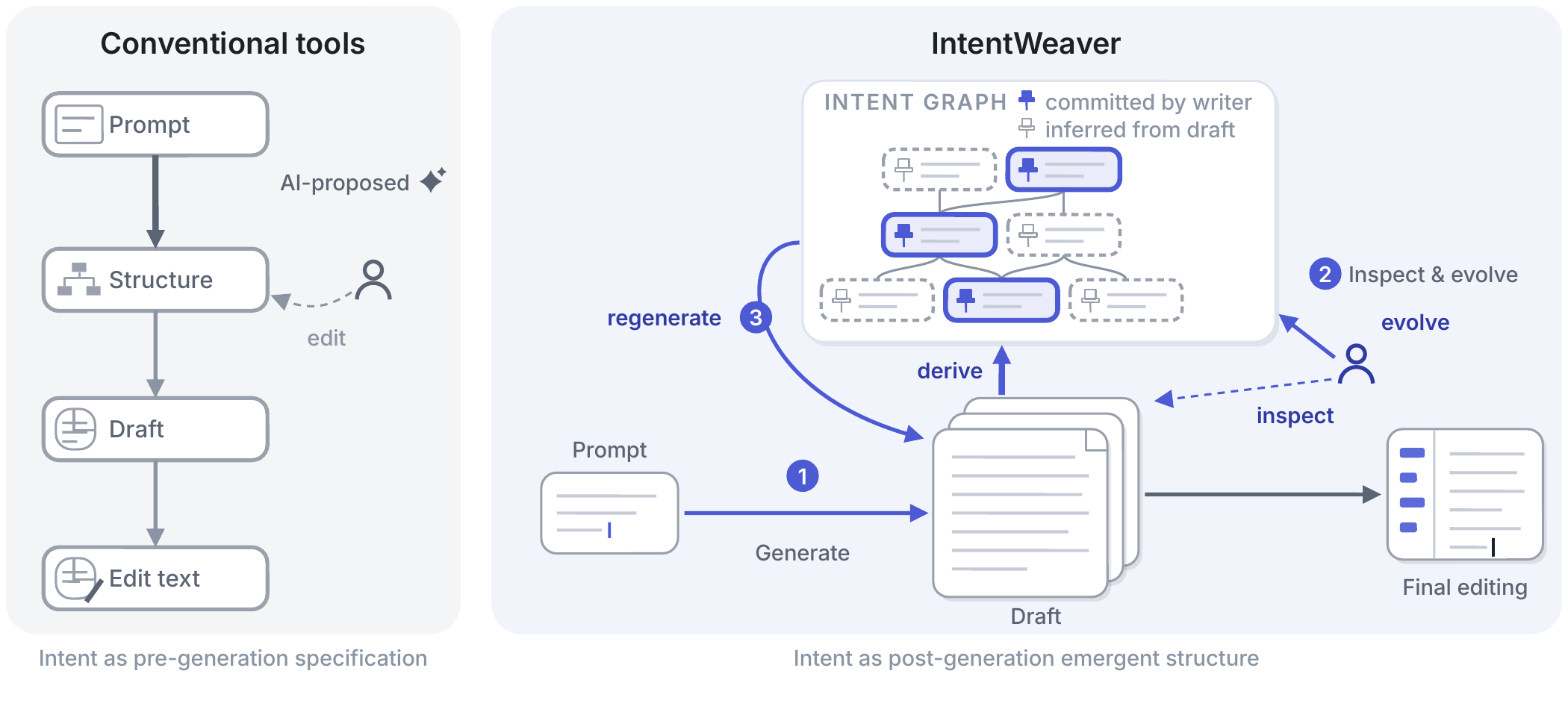}
  \caption{Conventional tools (left) treat intent as a specification before generation, so nothing registers when lower-level intents and texts drift from the higher-level ones they were meant to serve. IntentWeaver (right) derives the intent graph from each draft instead, keeping the whole hierarchy in view as it evolves: the writer generates (1), inspects the draft and evolves the graph (2), and regenerates (3), until the intents are well developed and the text fulfills them.}
  \Description{A two-panel diagram contrasting two workflows. The left panel, Conventional tools, is a downward chain: a prompt box, then a structure or outline box that the AI proposes and the writer edits, then a generated draft, then direct text editing; the panel is labelled intent as pre-generation specification. The right panel, IntentWeaver, is a loop: a prompt generates a draft in step one; an arrow labelled derive runs from the draft up to an intent graph whose cards are either committed by the writer, drawn solid with a filled pin, or inferred from the draft, drawn dashed with a hollow pin; in step two the writer inspects the draft and evolves the graph; in step three an arrow labelled regenerate runs from the graph back to a new draft; when graph and text agree the writer moves on to a final editing view. The panel is labelled intent as post-generation emergent structure.}
  \label{fig:planning-vs-exploration}
\end{teaserfigure}

\maketitle

\section{Introduction}

Human intent is fundamentally structured as a multi-level hierarchy, where lower-level intents dictate how an action is executed, while higher-level intents define why and to what end it is pursued \cite{Vallacher1987,kuutti1996activity}. 
In the context of human writing, this hierarchical organization manifests not as a static, upfront plan, but as an emergent phenomenon; intents continuously surface, sharpen, and dissolve as the writer engages with the evolving text \cite{flower1980cognition, flower1981cognitive}. 
Cognitive models of writing illustrate how authors construct and maintain these nested intent layers progressively through the act of composition, spanning from overarching conceptual goals (e.g., topic, audience, and genre), through section-level purposes and effects, down to specific content and lexical choices \cite{bereiter1987psychology, dimarco1993computational}. 
To prevent this highly fluid process from degenerating into undirected drift, writers must constantly govern coherence across different intent levels, 
keeping immediate textual outcomes aligned with their broader objectives. 
They achieve this through a recurrent \textit{``popping back up''} mechanism \cite{flower1981cognitive}, stepping back from local execution to re-evaluate progress against higher-level intents, reaffirming or revising these goals wherever appropriate, and propagating adjustments downward to reconcile drifted intents. 
Navigating this continuous loop of discovery and reconciliation is cognitively demanding, requiring metacognitive regulation to monitor, evaluate, and restructure goals across levels of abstraction in real time \cite{hacker_writing_2009}.

Large Language Models (LLMs) have catalyzed a new human-AI co-writing paradigm, shifting the writer's primary role from manual drafting to guiding and curating AI-generated text \cite{reza2025co, lee2024design}. This collaborative mode offers tangible cognitive benefits: AI outputs alleviate the ``blank page'' syndrome \cite{clark2018creative, gero2022sparks} by rapidly generating exploratory content, while confronting these concrete drafts helps writers recognize, crystallize, and refine nascent intents they might otherwise struggle to articulate \cite{gero2023social, lee2024design}.
However, this same generative capacity disrupts the natural cognitive dynamics of writing. Overwhelmed by the continuous influx of dense, AI-generated text, writers are forced into a reactive cycle of evaluating and editing local passages \cite{bhat2026reactive, siddiqui2025script}. Such micromanagement crowds out the metacognitive regulation \cite{lee2025impact} required to periodically ``pop back up'' and evaluate local text against higher-level goals \cite{bhat2026reactive, umarova2025problematic}.
Consequently, high-level intents stagnate while untethered lower-level intents proliferate and drift, leading to \textit{``intent incoherence''} -- a subtle but critical breakdown where writers feel highly productive editing AI outputs, yet invisibly decouple from or even violate their overarching purpose until the final text no longer reflects it \cite{bhat2026reactive, jakesch2023co}.

Recent human-computer interaction (HCI) research has attempted to mitigate intent-related challenges in co-writing with AI. One line of work externalizes intent through structured representations to guide AI generation, utilizing hierarchical goal decomposition (e.g., VISAR~\cite{zhang2023visar} and IntentFlow~\cite{kim2026intentflow}) or persistent intent specifications (e.g., Script\&Shift~\cite{siddiqui2025script}). Another line supports exploration of output alternatives given intents, by better organizing AI responses (e.g., Luminate~\cite{suh2024luminate}) or enabling rapid comparison of outcome variations (e.g., ABScribe~\cite{reza2024abscribe}). While these systems successfully assist in top-down formulation and articulation of intents, they tend to treat intent as a static, unidirectional constraint. 
There is a lack of support for the bottom-up process of writing: recognizing when local textual discoveries warrant the revision of a higher-level goal. 
Maintaining coherence across the intent hierarchy is thus largely left to the writer’s unaided initiative -- the very capacity that reactive co-writing erodes.

To address this gap, we conducted a \hyperref[sec:formative]{formative study} with 12 frequent AI co-writers, which revealed their frictions in higher-level intent monitoring, cross-level drift detection, and bidirectional intent reconciliation. Guided by these insights, we designed IntentWeaver, an interactive system that supports intent evolution and sustains coherence throughout the human-AI co-writing process.
Unlike prior systems that require upfront, structured intent externalization, IntentWeaver allows intent to emerge from, and evolve alongside, the AI-generated text.
Concretely, the system continuously derives and maintains an intent hierarchy from the evolving texts, surfaces coherence breakdowns and fulfillment gaps, and propagates user revisions at any level across the hierarchy.

To evaluate whether IntentWeaver mitigates intent incoherence and supports the evolution of high-level intents, and how writers appropriate and experience this system, we conducted a within-subjects study with 24 participants against a baseline system that preserves the core features of current mainstream AI co-writing tools. Results of mixed-methods analyses, combining task outputs, interaction logs, post-task questionnaires, and semi-structured interviews, showed that with IntentWeaver, participants returned to their high-level intents more often and evolved them further, and far less of their drift went unnoticed. Their final texts also realized the intents better, with no cost to agency or workload. We further identified three patterns of appropriation. Based on these findings, we offered design considerations for systems that maintain intent as a live structure, and consider where this approach applies beyond writing.

This work makes three key contributions: (1) IntentWeaver, a co-writing system that supports intent evolution and sustains intent coherence; (2) a mixed-methods evaluation of its effectiveness, usage patterns, and user experience; and (3) design considerations for treating intent as a live structure in human-AI collaboration.

\section{Related Work}
\subsection{Human-AI Co-Writing and Its Threat to Intent Coherence}
\label{sec: 2.1}

AI-powered writing assistance has transitioned from sentence-level suggestions and autocomplete \cite{clark2018creative, lee2022coauthor, yuan2022wordcraft} to a transformative ``co-writing'' paradigm \cite{reza2025co}. In this collaborative mode, writers offload large-scale text generation to LLMs and assume an editorial role over the model output. 
In a typical session, the writer describes what they need, the model returns a full draft, and further instructions refine it through successive rounds, as in tools such as Lex \cite{lex2024} and Gemini in Google Docs \cite{google2024gemini}.
Such a workflow now permeates professional communication and everyday genres such as social media posts \cite{reza2025co, lee2024design}. 
Co-writing offers clear cognitive benefits.
For writers who struggle to begin, a generated draft provides immediate material to work with, lowering the threshold from formulating language from scratch to editing readily available text \cite{clark2018creative, gero2022sparks}. 
For writers who know roughly what they want but cannot yet put it precisely, seeing the model's interpretation of their prompt can crystallize vague intentions into concrete choices \cite{gero2023social, lee2024design}.

However, this generative capacity may erode the epistemic benefit of writing, i.e., developing one's own understanding through composing \cite{umarova2025problematic}, by reshaping writers' cognitive process. The continuous influx of polished AI prose captures their immediate attention, shifting the core activity of composing from active ideation to reactive evaluation and copyediting \cite{bhat2026reactive, siddiqui2025script}. 
Trapped in prolonged, localized textual repair, writers rarely step back to check the unfolding text against what they themselves set out to achieve \cite{bhat2026reactive, umarova2025problematic}.
Because writers retain the ability to edit any portion of the text, they experience an illusion of complete control \cite{bhat2026reactive}, even as their arguments and opinions increasingly drift toward the model's suggestions \cite{jakesch2023co}. 
As a result, the finished text may reflect perspectives the author never intended to adopt, and these costs rarely surface in time to prompt correction \cite{bhat2026reactive, jakesch2023co}. 
This work aims to preserve the reflective value of offloaded generation while mitigating the induced reactive drift.
We propose to derive, structure, and maintain writer's evolving intents throughout the co-writing process.

\subsection{Writing Intent Hierarchy and Its Maintenance}
\label{sec: 2.2}

Writing is fundamentally a goal-directed, cognitive process where human intent is structured as a multi-level hierarchy \cite{flower1981cognitive}. According to the knowledge-transforming model \cite{bereiter1987psychology}, high-level conceptual goals dictate the why of the writing. Below them are two parallel execution layers: content space (deciding what to say within each part) \cite{bereiter1987psychology, grosz1995centering} and the rhetorical space (determining how to say it using specific stylistic and lexical choices) \cite{bereiter1987psychology, dimarco1993computational}.
This intent hierarchy is not a static plan fully created upfront; instead, it unfolds dynamically as the writer composes, evolving as the writer's understanding of the task matures \cite{flower1981cognitive}.

Because this hierarchy is highly fluid, writers need to actively maintain coherence across different levels while ensuring the text itself still fulfills those intents. They govern intent coherence by periodically ``popping back up'' from local, word-level editing to evaluate their progress against higher-level goals \cite{flower1981cognitive}. Ideally, this practice establishes a bidirectional metacognitive feedback loop. On the one hand, writers working on a local paragraph might discover a new angle, prompting them to revise their high-level intent and propagate that change downward to other sections, i.e., bottom-up discovery.
On the other hand, if a high-level intent remains valid, writers use it to pull drifting lower levels back into coherence, i.e., top-down reconciliation.

In practice, managing such a loop is cognitively demanding. 
Continuously shifting between concrete, word-level execution and abstract, high-level goals incurs context-switching costs that strain limited working memory \cite{monsell2003task, mccutchen1996capacity}, especially when both ends keep evolving. 
Moreover, humans naturally gravitate toward the immediate, low-friction feedback of local edits \cite{kool2010decision}, and a revision at one level affects intents and passages elsewhere that quickly exceed unaided memory \cite{cowan2001magical,flower1986detection}.
These demands are manageable in traditional writing: text grows slowly, every word is the writer's own, and change arrives in small increments. 
Co-writing with LLMs strips away these buffers and exacerbates the metacognitive vulnerabilities: the volume of AI prose amplifies the pull toward localized execution, leaving little cognitive bandwidth to pop back up \cite{bhat2026reactive}. 
Consequently, high-level intents stagnate while local AI-driven changes accumulate. 
Our work builds on this understanding of how intents are structured and maintained, aiming to sustain the coherence that co-writing erodes.

\subsection{Intent Communication in Human-AI Interaction}
\label{sec: 2.3}

In LLM-powered interactive systems, free-form prompts serve as the primary vehicle for communicating intent, yet users struggle to craft effective ones \cite{zamfirescu2023johnny}, as both envisioning the desired output and evaluating the model's actual delivery place substantial demands \cite{subramonyam2024envisioning}.
To mitigate this, HCI researchers across different domains increasingly treat intent as a manipulable artifact. 
Instead of forcing users to rely on raw text prompts, systems now let them inspect and adjust these intent artifacts to bridge the gap between human goals and AI output. 
Examples include hierarchies of design semantics in user interface development \cite{park2026bridging}, code blueprints in programming \cite{liu2023wants}, visual prompt explorers in text-to-image creation \cite{brade2023promptify}, and motion sketches in animation generation \cite{li2026sketchdynamics}.

AI-assisted writing systems have followed a similar path \cite{zhao2025making}, generally using one of two approaches. The first externalizes intent as a structured representation to guide generation. For instance, VISAR renders argumentation goals as a visual hierarchy and lets writers preview each goal via a rapidly prototyped draft \cite{zhang2023visar}. IntentFlow extracts intents from the prompts into editable components, visually linking them to the corresponding output segments \cite{kim2026intentflow}. Script\&Shift maintains persistent intent declarations that contextualize AI assistance throughout composing \cite{siddiqui2025script}.
However, in these systems, intent only flows downward. They do not support the bottom-up process where writers discover new ideas while editing the drafts and need to change their original high-level goals. 
The second approach focuses on output exploration, organizing alternative LLM responses along generated dimensions \cite{suh2024luminate} or enabling rapid comparison of variations \cite{reza2024abscribe}.
A critical gap in both approaches is that no mechanism keeps the intent representation in step with the text it once produced. 
As the writer keeps editing, these tools do not track how new revisions and emerging intents affect the original goals, nor detect when lower-level intents drift from the higher-level ones they once served. 
The burden of maintaining coherence falls entirely on the writers at the exact moment their metacognitive capacity is most taxed by reactive copyediting, so that resulting documents silently betray their original intents.

To bridge this gap, we take inspiration from goal-oriented requirements engineering \cite{van2001goal, amyot2010evaluating}, a field in software engineering that maps how intentional elements impact one another and computes how a change in one ripples through the whole system. 
We adapt this logic to writing and build a system that derives the intent hierarchy from the evolving draft, evaluates it, and propagates revisions across levels, so that a writer's goals and text evolve together.

\section{Formative Study}
\label{sec:formative}

To understand how writers lose coherence across intents when co-writing with LLMs, we conducted a formative study; its findings, together with prior work, yielded three design requirements that guided IntentWeaver.

\subsection{Study Procedure}

We recruited twelve participants under an IRB-approved protocol, reaching them through social media posts and personal referrals (seven female, five male; PF1--PF12).
All participants regularly co-write with AI (seven several times a day, four once or twice a day, one several times a week) across academic, professional, and everyday genres, and reported having experienced episodes of becoming absorbed in refining local details of AI output while losing sight of the piece's overall direction.
In semi-structured interviews, participants reconstructed recent co-writing sessions step by step, describing their interactions with AI, the thinking behind each move, and the difficulties they encountered along the way. 
We also asked them to recall the same dynamics when writing on their own, using the contrast to surface what co-writing with AI disrupts, and solicited suggestions for potential support. 
Each session lasted around 40 minutes, compensated at \$10. 
Interviews were audio-recorded, transcribed, and subjected to inductive thematic analysis by two researchers who developed the codebook through iterative discussion \cite{hsieh2005three, braun2006using}.
The codebook stabilized by the ninth interview; the final three interviews yielded no new themes, suggesting saturation \cite{saunders2018saturation}.

\subsection{Summary of Findings}
\label{sec:formative-findings}

Three themes trace how coherence maintenance breaks down: attention is pulled into local work, the divergence widens without notice, and by the time it is noticed the way back has grown prohibitively expensive.

\subsubsection{Pulled into Local Work}

The pull began with the arrival of the text itself. Because AI output came back fluent, well organized, and correct sounding (PF1--6, PF9--11), participants described an \textbf{illusion of completion}. 
PF8 recalled that the output \textit{``feels nearly finished—I'd only need to fix details.''} 
Participants accordingly slid from author into proofreader, their evaluation narrowing from assessing the direction to scanning its surface. 
This proofreading stance was then reinforced by an \textbf{asymmetry in evaluation cost}. 
Verifying a phrase takes seconds and closes cleanly, whereas judging whether the piece still serves one's higher-level intents means rebuilding those intents and rereading the whole; participants rarely had energy left for the latter (PF1, PF10), and admitted to \textit{``instinctively satisfying whatever was easiest to evaluate''} (PF7). 
Meanwhile, rapid generation surfaced masses of eye-catching detail demanding polish, such as awkward phrasings and rough joints (PF3, PF6--8). 
These imperfections acted as reading friction, while higher-level intents offered no comparable pull on attention (PF12). 
Participants thus stopped to fix even details they expected to delete anyway, partly out of perfectionism (PF8), but more because the friction made it hard to read on (PF6--8), a \textbf{compulsive cleanup of surface imperfections}. 
Prior work documents that AI output captures writers' attention~\cite{bhat2026reactive}; our participants explain why the capture persists. 
The pull from the text turned out to be a rational allocation: the cheapest evaluations, and the loudest signals, all lie at the bottom.

\subsubsection{Drifting without Notice}

Being absorbed in local work does not by itself set the text drifting from what the writer intended. 
Three patterns did, each keeping the divergence out of sight.
Much of the time, participants' \textbf{higher-level intents were absent from the interaction}. 
Rather than working out what they wanted at a deeper level, participants habitually reduced it to concrete, ready-to-execute instructions for the AI: PF12 described turning a global framing goal into \textit{``one specific sentence to add.''} 
Once the AI fulfilled the instruction, participants took the goal behind it as settled and stopped attending to it (PF12). 
And with the goal itself stated nowhere, later text could drift from it with nothing to check against.
Divergence also escaped notice turn by turn. 
Within a single round, the output normally strayed from the writer's intents only slightly (PF2, PF9), too little to cross the threshold that triggers alarm (PF3, PF7).
Across rounds, these \textbf{micro-drifts accumulated} into a displacement that no single moment had ever flagged.
Drift could also be invited outright. AI output kept introducing points that participants found valuable yet outside the task and plan (PF1, PF9). 
The reaction, as PF4 put it, was \textit{``this is good, let's write it this way,''} and participants turned the piece then and there without asking what broader intent the new point represented or what its adoption meant for existing ones. These were \textbf{novelties adopted without consolidation}.
With this rethinking skipped, participants' higher-level intents never matured to reflect the new direction, leaving later local choices unguided, and they \textit{``kept flip-flopping without a clear line''} (PF3, PF9).

\subsubsection{The Costly Way Back}

Participants did notice, at times. 
But when they tried to reconcile their intents, the attempt failed in both directions, a difficulty rooted in \textbf{ripple effects across interdependent intents}. 
When they asked for the correction at the high level, it foundered on invisible dependencies (PF4--7). As PF3 explained, \textit{``When you adjust something high-level, it does not know which lower-level intents should change along with it [...] the result feels forced and disjointed.''}
When they instead revised the text locally, each change disturbed what had been built around the original: the reworked part no longer fit the surrounding logic and transitions, and repairing those pulled in still more of the text (PF6, PF8, PF10). 
Tracing the dependencies among their intents, and how each change would shift them, was more than participants could manage (PF4); 
PF6 described ending up issuing feedback that contradicted his own earlier intents and \textit{``spinning in place,''}
until nothing short of regenerating the whole text could reconcile the intents (PF6, PF10). 
But regeneration meant forfeiting \textbf{invested thought and craft anchored to the current output}.
Polished passages carried sunk cost \cite{arkes1985psychology} (PF3, PF6), and all the thinking participants had built up along the way was tethered to this specific output; neither of which they could bring themselves to discard (PF3, PF9--11).
PF2's wish for a system that \textit{``preserves the reasoning behind past edits''} reveals exactly what regeneration currently destroys. 
With no workable way out, a repair that fell apart and a reset unaffordable, participants could only stay fixated on the text they had.

\subsection{Design Requirements}
\label{sec:formative-drs}

\begin{figure*}[t]
  \centering
  \includegraphics[width=\textwidth]{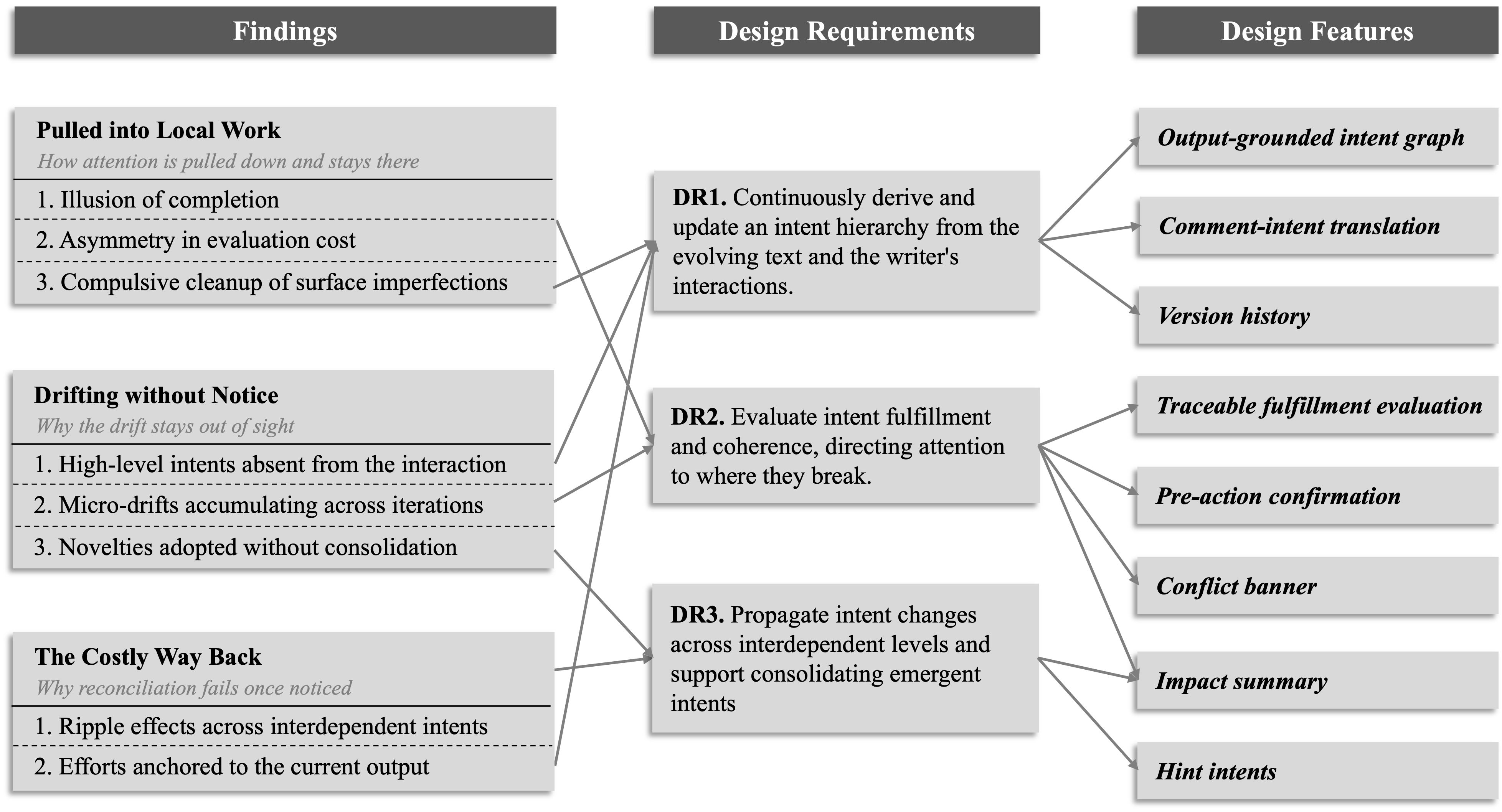}
  \caption{Mapping from formative findings to design requirements and system features. The three themes (left) motivate DR1--DR3 (center), which in turn are realized by the features of IntentWeaver (right).}
  \Description{A three-column mapping diagram. The left column, Findings, holds three themes with their sub-findings. Pulled into Local Work, how attention is pulled down and stays there: illusion of completion; asymmetry in evaluation cost; compulsive cleanup of surface imperfections. Drifting without Notice, why the drift stays out of sight: high-level intents absent from the interaction; micro-drifts accumulating across iterations; novelties adopted without consolidation. The Costly Way Back, why reconciliation fails once noticed: ripple effects across interdependent intents; efforts anchored to the current output. Arrows fan from these sub-findings, crossing between themes, into the middle column of three design requirements. DR1, continuously derive and update an intent hierarchy from the evolving text and the writer's interactions. DR2, evaluate intent fulfillment and coherence, directing attention to where they break. DR3, propagate intent changes across interdependent levels and support consolidating emergent intents. Arrows then fan from each requirement to the right column of eight design features: output-grounded intent graph, comment-intent translation and version history from DR1; traceable fulfillment evaluation, pre-action confirmation and conflict banner from DR2; impact summary and hint intents from DR3.}
  \label{fig:formative-pipeline}
\end{figure*}

These findings converge on three design requirements (Figure~\ref{fig:formative-pipeline}).

\textbf{DR1: Continuously derive and maintain an intent hierarchy from the evolving text and the writer's interactions.}
Writers' higher-level intents require an externalized manifestation, yet it is difficult, if not impossible, to solicit a complete, finalized one upfront. More reliable sources of a writer's latest thinking are the evolving text and their ongoing engagement with it; every comment or revision on the text reveals something about what the writer wants. The system should therefore continuously transform the texts and interactions into a hierarchical intent representation and maintain it dynamically across model regenerations. The thought and effort that once perished with a disposable draft thereby settle into a persistent, inspectable schema that outlives any single textual output.

\textbf{DR2: Evaluate intent fulfillment and coherence, directing attention to where they break.}
An explicit hierarchy alone does not restore vigilance. The system should take over the analytical appraisal that writers see no need for and cannot afford during cognitively demanding co-writing tasks. Upon every local or global change, it should evaluate how well the draft fulfills each intent and whether different levels of the hierarchy still cohere. The assessment outcome should be surfaced through non-intrusive visual indicators in real time, flagging subtle, per-turn deviations before they compound into invisible drifts. This provides a check independent of the writer’s own initiative, prompting metacognitive reflection on the text's actual trajectory.

\textbf{DR3: Propagate intent changes across interdependent levels and support consolidating emergent intents.}
The intent hierarchy is in constant motion, and an adjustment at one level inevitably impacts others. These nested dependencies are often difficult to trace, forcing writers to choose between disjointed patches and total rewrites. The system should thus automatically propagate the consequences of any intent change, including emergent ones introduced by the AI co-writer, 
both upward and downward to surface the resulting semantic shifts, preventing them from silently hijacking the text's direction. The system should consolidate such evolution and its implications into the hierarchy and remain human-in-the-loop, allowing writers to review and approve suggested cascade adjustments.

\section{Design}
\label{sec:system}
We introduce IntentWeaver, an AI co-writing system that inverts the traditional relationship between intent and text (Figure~\ref{fig:planning-vs-exploration}). Conventional AI-assisted writing tools treat intent as a rigid, pre-generation specification, prompting the writer to comprehensively develop a goal structure before the AI drafts any text. IntentWeaver, by contrast, treats intent as an emergent, post-generation structure. It positions the AI-generated draft not as a final product, but as an inspectable, reflective medium through which the writer can discover, calibrate, and refine their goals.

To this end, a writing session in IntentWeaver progresses through an iterative, three-stage loop: 1) The writer states what they want; the system generates a read-only draft and, alongside it, automatically derives a hierarchy of writing intents from that text and visualizes it. 2) The writer inspects the draft and, informed by what it reveals, directly manipulates and evolves the visualized intent hierarchy by adding, deleting, or refining goals. 3) When the active draft has exhausted its utility for intent development, the writer triggers a regeneration; the system synthesizes a new draft tailored to the updated intent hierarchy, starting the next collaborative loop. This loop repeats until the writer's intents are well-developed and fulfilled by the text. At this point, the writer transitions to a final editing mode, unlocking direct text editing for surface-level polishing and exporting the final document.

The rest of this section covers a usage scenario (Section~\ref{sec:scenario}), the theoretical basis (Section~\ref{sec:hierarchy}), how the intent graph is derived and presented (Section~\ref{sec:graph}), and how writers steer and evolve it (Section~\ref{sec:steering}); implementation details, propagation algorithms, and model prompts are in the supplementary material.

\begin{figure*}[p]
  \centering
  \begin{subfigure}{\textwidth}
    \centering
    \includegraphics[width=\textwidth]{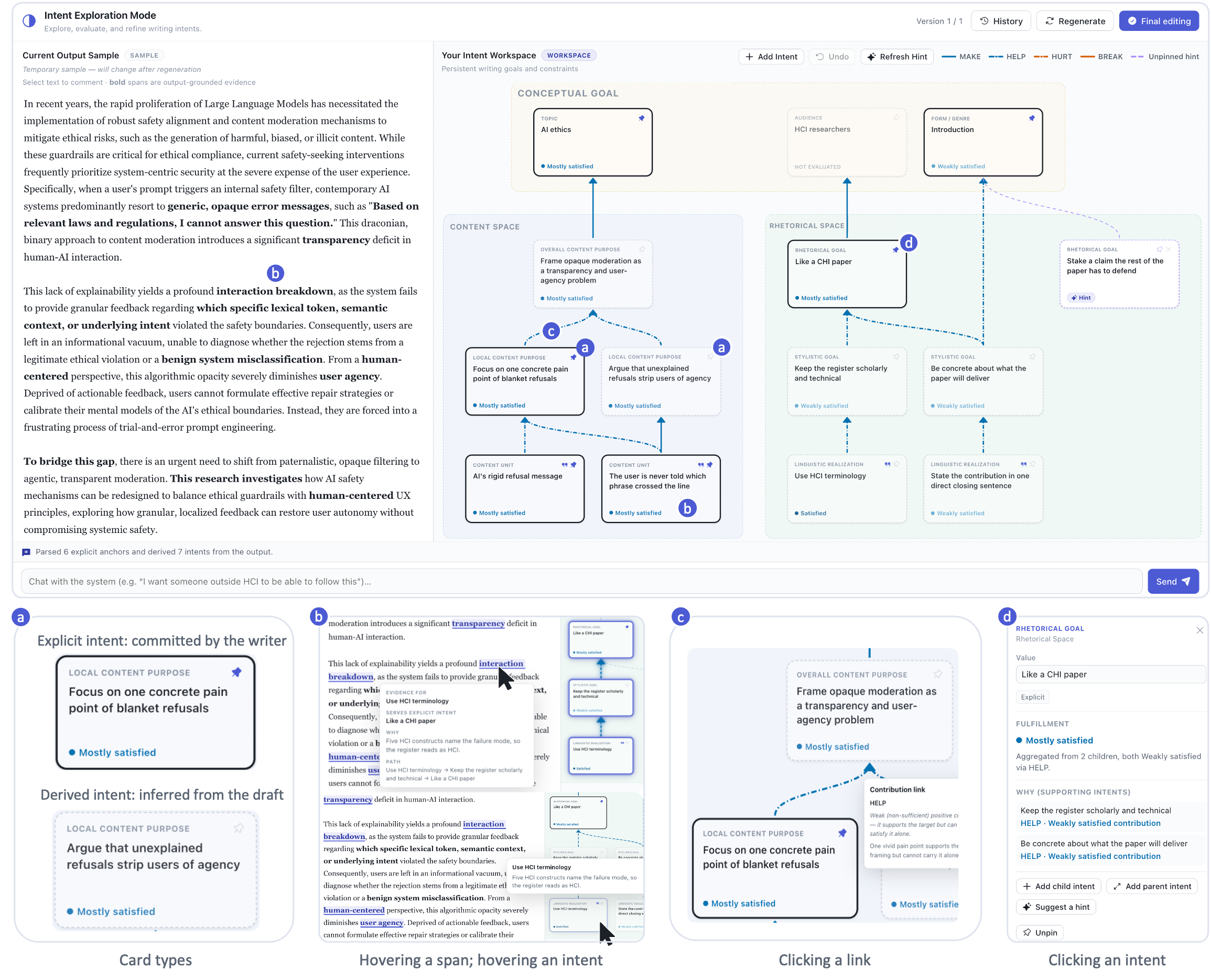}
    \caption{The interface: the read-only draft (left) and the intent graph (right). Callouts a--d show the inspection interactions.}
    \label{fig:interface}
  \end{subfigure}

  \vspace{1pt}

  \begin{subfigure}{\textwidth}
    \centering
    \includegraphics[width=0.97\textwidth]{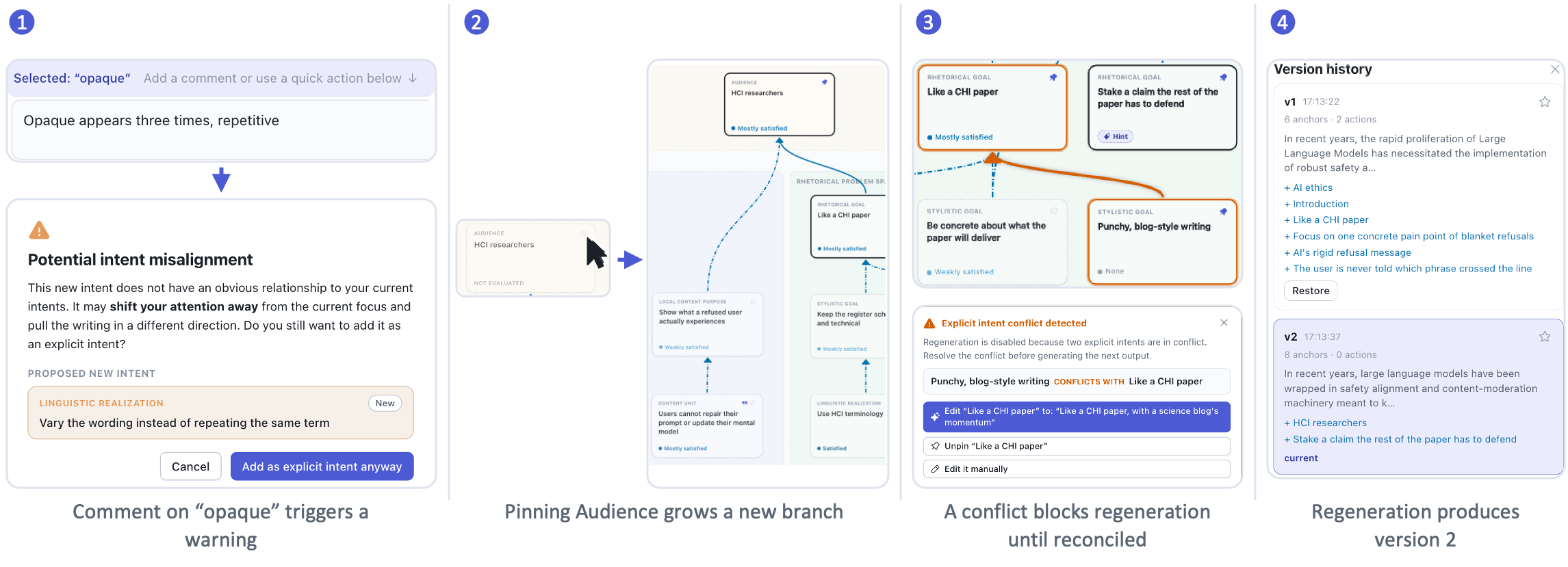}
    \caption{Four moments from the usage scenario in Section~\ref{sec:scenario}.}
    \label{fig:storyboard}
  \end{subfigure}

  \caption{IntentWeaver: the interface with its inspection interactions (a) and the usage scenario (b).}
  \Description{Two stacked panels. Panel a is a screenshot of the IntentWeaver interface in intent exploration mode. Its left pane holds a read-only draft about opaque AI content moderation, with bold spans marking the text that grounds bottom-level intents and a comment composer beneath. Its right pane is the intent workspace, split into three bands: conceptual goal, holding topic, audience and form or genre cards; content space, holding overall content purpose, local content purpose and content unit cards; and rhetorical space, holding rhetorical goal, stylistic goal and linguistic realization cards. Solid cards with a filled pin are explicit intents, dashed cards with a hollow pin are derived intents, and every card carries a fulfillment label ranging from satisfied to not evaluated. Links between cards are colored and dashed by contribution type, with a legend for make, help, hurt, break and unpinned hint. Four enlarged callouts show the inspection interactions: a, the two card types side by side; b, hovering a bold span to reveal a tooltip naming the evidence, the explicit intent it serves, why, and the path between them; c, clicking a link to read the rationale for its contribution type; and d, clicking an intent to open a detail panel with its value, its fulfillment, the supporting intents, and actions to add a child or parent, suggest a hint, or unpin. Panel b is a four-frame storyboard of the usage scenario. Frame one, a comment on the word opaque is translated into a proposed linguistic realization and triggers a potential intent misalignment warning that the writer can cancel or override. Frame two, pinning the derived audience card grows a new branch of intents in the graph. Frame three, a conflict banner reports that punchy, blog-style writing conflicts with like a CHI paper, disables regeneration, and offers three ways out: accept a synthesized compromise, unpin one side, or edit by hand. Frame four, the version history panel lists version one and version two with their anchors, actions and added intents, and a restore button.}
  \label{fig:usage-scenario}
\end{figure*}

\subsection{Usage Scenario}
\label{sec:scenario}

To illustrate, we follow Wei, a researcher drafting the introduction of a paper. She describes her task in a prompt: an introduction about AI ethics that reads ``like a CHI paper'' and highlights a pain point, that a blocked prompt returns a rigid refusal without naming the phrase that triggered it. 
The system returns a read-only draft alongside an intent hierarchy (Figure~\ref{fig:interface}). User-declared intents (bold-bordered cards) come directly from Wei's prompt, including the topic, genre, CHI-paper style, and the pain point. Inferred intents (dashed cards) are latent goals the LLM infers the draft is pursuing, such as applying HCI terminology.

Wei notices that her high-level \card{Introduction} genre intent is only weakly fulfilled. The chain beneath it reveals that the ending lacks a clear statement of contribution. On her way to inspect it she is distracted: the word ``opaque'' keeps recurring. Because the draft is read-only under exploration mode, she selects the word and adds a comment requesting more diverse wording. 
The system alerts Wei that this minor stylistic goal does not serve any of her objectives and risks derailing her attention (Figure~\ref{fig:storyboard} (1)). Wei cancels, recognizing that the repetition may resolve in the next rewrite.

Back in the top band she notices a dashed card, \card{Audience: HCI researchers}, that the system inferred but never asked her about. Recognizing that this could help elicit the paper's contribution, she ``pins'' it to transform this background context into a commitment. The system propagates this change downward by re-evaluating the draft through this newly prioritized audience lens and automatically growing a new branch of support anchored in textual evidence (Figure~\ref{fig:storyboard} (2)). The UI displays an impact summary detailing these updates.

To help resolve the weak ending of \card{Introduction}, IntentWeaver displays a suggested ``hint'' card \card{stake a claim the rest of the paper has to defend} beside it. Wei pins this intent and the card turns bold, making it an active goal for the next iteration. 
Reading the stiff prose again, she comments that it could be punchier, more like a blog post. The system proposes the intent but warns that this style directly opposes Wei's initial commitment to write \card{like a CHI paper}. Wei sticks to her opinion, resulting in two of her committed intents in conflict. The system temporarily disables regeneration to prevent structural fragmentation. To clear the bottleneck, the conflict banner proposes an evolved, synthesized compromise: \card{like a CHI paper, with a science blog's momentum} (Figure~\ref{fig:storyboard} (3)).
Wei accepts; the conflict clears, coherence is restored across the hierarchy, and Regenerate unlocks.

She regenerates, and the hierarchy is rebuilt from the new draft with fresh derived intents (Figure~\ref{fig:storyboard} (4)): \card{Introduction} is now satisfied, the ``blog-style momentum'' is realized through shorter, more active sentences, and the repetitive vocabulary issue is gone with the old draft.
After two more rounds the hierarchy and text are in sync. Wei unlocks final editing to perform manual polishing on a text whose underlying conceptual integrity she fully trusts.

\subsection{Theoretical Basis and Representation Model}
\label{sec:hierarchy}

\begin{figure*}[t]
    \centering
    \includegraphics[width=0.75\textwidth]{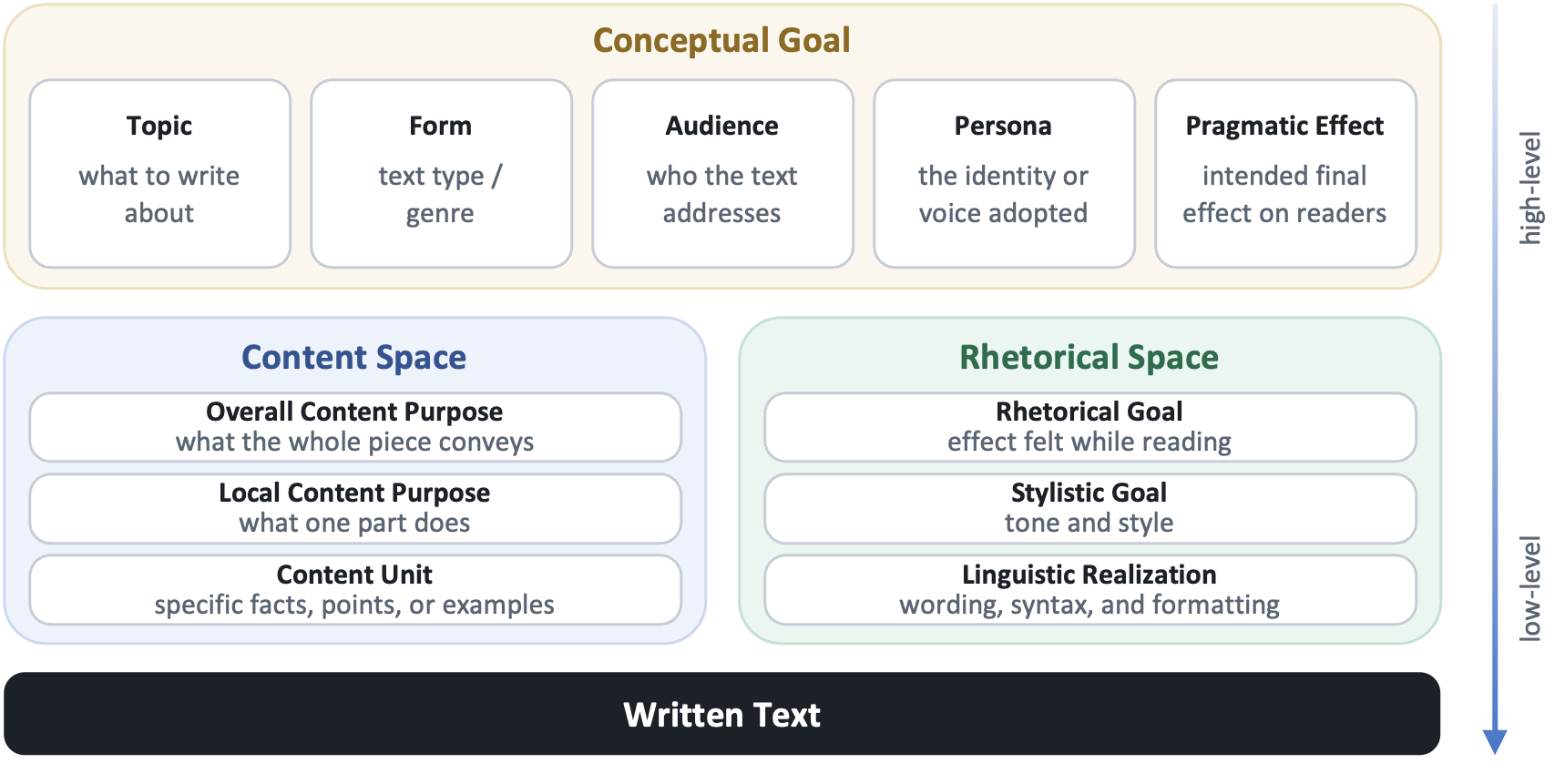}
    \caption{The writing intent hierarchy synthesized from cognitive writing theories. Conceptual goals sit above the content and rhetorical spaces, which decompose in parallel into progressively more local intents, ultimately realized in the written text. Detailed definitions and examples of each element are provided in the supplementary material.}
    \Description{A diagram of the writing intent hierarchy. A top band, conceptual goal, holds five elements: topic, what to write about; form, text type or genre; audience, who the text addresses; persona, the identity or voice adopted; and pragmatic effect, the intended final effect on readers. Below it sit two parallel bands. The content space holds overall content purpose, what the whole piece conveys; local content purpose, what one part does; and content unit, specific facts, points or examples. The rhetorical space holds rhetorical goal, the effect felt while reading; stylistic goal, tone and style; and linguistic realization, wording, syntax and formatting. A dark band at the bottom is the written text. An arrow down the right edge runs from high-level to low-level.}
    \label{fig:writing-intent-framework}
  \end{figure*}

IntentWeaver synthesizes cognitive theories of writing (Section \ref{sec: 2.2}) into a unified, hierarchical framework (Figure~\ref{fig:writing-intent-framework}), which structures writing as a high-level conceptual goal space directing two parallel spaces: content and rhetoric. At the conceptual apex, we map the recurring dimensions along which writers initially frame their task: topic, form, audience, persona, and pragmatic effects~\cite{flower1980cognition, dimarco1993computational}. Within the content space, the overall semantic purpose is decomposed hierarchically into the local parts~\cite{grosz1995centering}. Within the rhetorical space, communicative goals break down into stylistic objectives, ultimately realized through concrete linguistic choices~\cite{dimarco1993computational}.

For the relations among intents and their evaluation logic, we adapt the formalisms of goal-oriented requirements engineering~\cite{van2001goal, amyot2010evaluating}. Directed links connect lower-level intents (children) to the higher-level intents (parents) they support; each link carries one of four qualitative contribution types: \lvl{make} (sufficient positive), \lvl{help} (partial positive), \lvl{hurt} (partial negative) and \lvl{break} (sufficient negative). An intent's fulfillment level is represented on a qualitative scale ranging from \lvl{satisfied} to \lvl{denied}. For intermediate and high-level nodes, this value is computationally aggregated from its children: each incoming link contributes a weighted influence determined jointly by the child's individual fulfillment state and the link's contribution type. For example, under a \lvl{make} relationship, a single \lvl{satisfied} child node is sufficient to render its parent node as \lvl{satisfied}. Bottom-level nodes are evaluated by an LLM directly against the text spans they are grounded in.
This structured abstraction lets IntentWeaver trace dependencies, detect incoherence, and systematically propagate the consequences of a change across levels, as the next two subsections detail.

\subsection{Deriving and Presenting the Intent Graph} 
\label{sec:graph}

\subsubsection{Derivation from the Draft (DR1, DR2)}
\begin{figure*}[t]
  \centering
  \includegraphics[width=\textwidth]{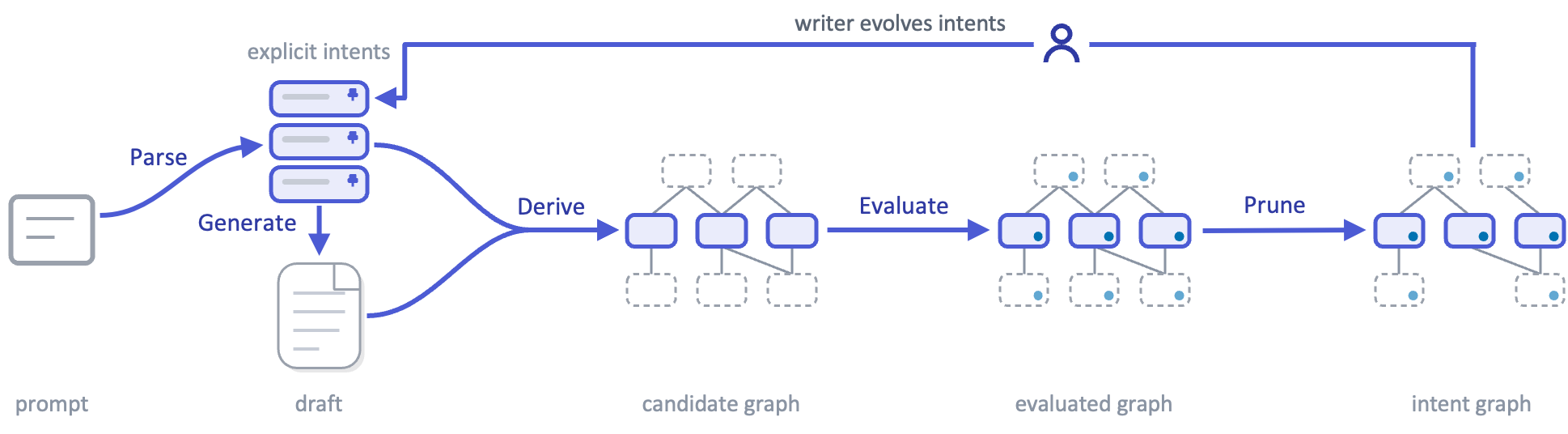}
  \caption{How IntentWeaver derives the intent graph. The prompt is parsed into explicit intents, from which a draft is generated. Explicit intents and draft together yield a candidate graph, which is evaluated bottom-up from grounded spans and pruned to intents relevant to the explicit ones. The writer's later actions evolve the explicit intents, from which the next round derives.}
  \Description{A left-to-right pipeline diagram. A prompt box is parsed into a stack of three explicit intent cards, each carrying a filled pin. An arrow labelled generate leads from those cards down to a draft page. Arrows from the explicit intents and from the draft converge into one arrow labelled derive, which produces a candidate graph of solid cards for explicit intents and dashed cards for derived ones across three levels. An arrow labelled evaluate produces the same graph with a fulfillment dot on every node. An arrow labelled prune produces the final intent graph with two dashed nodes removed. A return arrow along the top, carrying a writer icon and the label writer evolves intents, runs from the intent graph back to the explicit intents.}
  \label{fig:pipeline}
\end{figure*}

Figure~\ref{fig:pipeline} shows how IntentWeaver derives the intent graph. The writer's initial prompt is first parsed into \emph{explicit intents}, each typed by an element of the hierarchy, and a draft is generated from them. Explicit intents are those the writer has committed to, whether in the prompt or through later interaction; they persist across regenerations as the anchors from which every new draft is generated. From these anchors and the draft, IntentWeaver infers \emph{derived intents}: intents the writer never stated but which the system reads the current draft as pursuing around each explicit intent, downward to the content units and phrasings that realize it and upward to the broader purpose it plausibly serves. It also infers the links among all intents, each typed by its contribution. Unlike explicit intents, derived intents are transient and may change or disappear with the next draft. With the candidate graph in place, evaluation proceeds bottom-up. Only bottom-level intents, content units and linguistic realizations, are grounded to verbatim spans of the draft as evidence and assessed directly for fulfillment; every higher level is computed from them by propagation. The graph is then pruned to remove derived intents of little relevance to the explicit ones.

This derivation is deliberately anchored on explicit intents rather than exhaustive. A draft admits any number of rhetorical choices and content claims, and enumerating them would bury the intents that matter under those that merely happen to be present, and, by making every dimension of the text visible, would invite the writer to attend to dimensions they never cared about. That is the pull toward local work in another guise. IntentWeaver therefore retains only derived intents with sufficient relevance to the explicit ones. 
The graph is thus not a comprehensive analysis of the text but a lens for understanding and evolving one's intents against the output.

\subsubsection{Visual Encodings (DR2)}
As shown in Figure~\ref{fig:interface}, IntentWeaver's interface has two panes. 
The left pane holds the current draft, in which spans that ground bottom-level intents are set in bold, so that the writer's eye is drawn first to the text most relevant to their intents. 
The right pane is the intent workspace, partitioned into the three spaces of the hierarchy. 
Each intent is a card labeled with its element type and value, with fulfillment level shown as a colored label at the bottom. 
Explicit intents have a solid, heavier border and a filled pin; derived intents a lighter, dashed border and a hollow pin. Links run from lower to higher intents and follow the legend in the toolbar: blue for positive contributions and orange for negative, solid for sufficient strength (\lvl{make}, \lvl{break}) and dash-dot for partial (\lvl{help}, \lvl{hurt}). 
A hint intent, described later, appears as a dashed purple card tethered to the explicit intent it would serve.

\subsubsection{Inspection Interactions (DR2)}
Beyond automating evaluation, IntentWeaver lets the writer trace how it arrived at each judgment. 
For any intent above the bottom level, hovering or clicking the card reveals which lower-level intents contribute to it, and by how much. 
For a bottom-level intent, hovering it highlights the spans it is grounded in on the left (Figure~\ref{fig:interface} (b)). 
The tie runs both ways. 
Hovering a bold span reveals the bottom-level intent it evidences and the explicit intent it ultimately supports, explains in a line what the span does for them, and lights up the path between them in the graph. 
The writer can also click any link to read the rationale behind the contribution type it carries (Figure~\ref{fig:interface} (c)). 
These interactions let the writer audit the system's reading of the draft at every level before acting on it.

\subsection{Steering and Evolving Intents}
\label{sec:steering}

\subsubsection{Acting on the Graph (DR1)}
The writer cannot edit the draft directly.
Every action applies to the intent graph instead, and the draft changes only when the writer regenerates it from the updated intents.
Closing the editing channel while keeping the intent channel open shuts off the move our participants found hardest to resist, fixing surface imperfections in place, and routes whatever the writer wants to change into the hierarchy, where it outlives the draft.

The writer can steer the graph in two ways. The more direct way is to act on cards directly. 
Realizing that a derived intent names something they do want, the writer pins it, making it explicit; finding that an explicit intent no longer matters, they unpin it, letting it lapse at the next generation. 
When the draft sharpens their sense of what an intent should be, they can edit the value; 
when it surfaces something no card yet captures, they can add a new intent as a child or parent of an existing one or on its own.
The other way is the familiar gesture of co-writing. 
Beneath the draft sits a composer; the writer can select a passage and comment on it there, or type without a selection to comment on the draft as a whole. 
IntentWeaver translates each comment into a candidate action of the first kind.

\subsubsection{Confirming Actions (DR2)}
An action never takes effect on its own. Before committing it, the system checks where the candidate would attach to the current graph and how it would contribute, and presents this in a confirmation dialog, so the writer sees what the action will do to their intents. 
When the candidate relates to none of the existing intents or works against one, the dialog carries a warning that says so (Figure~\ref{fig:storyboard} (1)). These are the two ways incoherence announces itself.
The writer may proceed regardless. The warning asks only for a moment of reflection on whether the concern deserves a place among their intents. 
In our study, this is precisely the moment at which local concerns slipped in unexamined. 
An action committed over a warning is one the writer has weighed and chosen to stand by, marking deliberate exploration rather than blind drift even when the new intent serves nothing else they currently want.

\subsubsection{Local Update and Impact Summary (DR1, DR2, DR3)}
Committed actions update the graph locally, touching only what the action could have changed (Figure~\ref{fig:local-update}). 
Where the changed intent lacks a reading, the system derives one from the draft in whichever direction is missing. Downward, this yields a path of derived intents that realize it, ending in bottom-level ones grounded in the text (Figure~\ref{fig:storyboard} (2)); upward, the conceptual goals it serves; or both at once. 
Where an intent's value changes, the system re-judges its links, which may turn a contribution from positive to negative or back. 
Fulfillment is then re-propagated through the affected region, and an impact summary lists what the action did: which intents appeared or changed status, which links changed type, and how far the fulfillment of each affected intent moved, with a one-step undo.

\newcolumntype{L}[1]{>{\raggedright\arraybackslash}p{#1}}

\begin{figure*}[t]
\centering
\setlength{\tabcolsep}{4pt}
\setlength{\luU}{\dimexpr\textwidth-10\tabcolsep\relax}
\setlength{\luA}{0.13\luU}\setlength{\luB}{0.21\luU}\setlength{\luC}{0.18\luU}\setlength{\luD}{0.24\luU}\setlength{\luE}{0.24\luU}
\setlength{\luBC}{\dimexpr\luB+\luC+2\tabcolsep\relax}          
\setlength{\luXb}{\dimexpr\luA+2\tabcolsep\relax}                 
\setlength{\luXc}{\dimexpr\luXb+\luB+2\tabcolsep\relax}
\setlength{\luXd}{\dimexpr\luXc+\luC+2\tabcolsep\relax}
\setlength{\luXe}{\dimexpr\luXd+\luD+2\tabcolsep\relax}
\setlength{\luH}{124pt}               
\footnotesize
\begin{tabular}{p{\luA} p{\luB} p{\luC} p{\luD} p{\luE}}
\begin{tikzpicture}[baseline=(f.north)]
  \node[lupanel,text width=\dimexpr\luA-8pt\relax,minimum height=\luH] (f) {%
    \lutype{comment}\\[-1pt]``Sentences too wordy right now''\\[3pt]
    \lutype{translated to}\\[-1pt]add \card{keep sentences short}, a stylistic goal};
  \lubadge{f}{A}
\end{tikzpicture} &
\multicolumn{2}{p{\luBC}}{%
\begin{tikzpicture}[baseline=(f.north)]

  \node[
    draw=luwarn!55,
    fill=white,
    rounded corners=5pt,
    line width=0.5pt,
    minimum width=\luBC,
    minimum height=\luH,
    inner sep=0pt
  ] (f) {};

  \coordinate (warning) at
    ([xshift=12pt,yshift=-10pt]f.north west);

  \path[
    fill=luwarn,
    rounded corners=0.6pt
  ]
    ([xshift=-4.5pt,yshift=-3.5pt]warning) --
    ([xshift=4.5pt,yshift=-3.5pt]warning) --
    ([yshift=4.5pt]warning) -- cycle;

  \node[
    text=white,
    font=\bfseries\tiny
  ] at ([yshift=-0.8pt]warning) {!};

  \node[
    anchor=north west,
    text width=\dimexpr\luBC-14pt\relax,
    align=left,
    inner sep=0pt,
    font=\bfseries\footnotesize
  ] (title) at
    ([xshift=7pt,yshift=-20pt]f.north west)
    {Potential intent misalignment};

  \node[
    anchor=north west,
    text width=\dimexpr\luBC-14pt\relax,
    align=left,
    inner sep=0pt,
    font=\scriptsize
  ] (message) at
    ([yshift=-3pt]title.south west)
    {This new intent does not have an obvious relationship to your current
    intents. It \textbf{may shift your attention away} from the current focus
    and pull the writing in a different direction. Do you still want to add
    it as an explicit intent?};

  \node[
    anchor=north west,
    inner sep=0pt,
    font=\bfseries\tiny,
    text=luink!55
  ] (proposed-label) at
    ([yshift=-4pt]message.south west)
    {PROPOSED NEW INTENT};

  \node[
    anchor=north west,
    draw=luwarn!50,
    fill=luwarm,
    rounded corners=3pt,
    line width=0.4pt,
    minimum width=\dimexpr\luBC-14pt\relax,
    minimum height=22pt,
    inner sep=0pt
  ] (proposal) at
    ([yshift=-2pt]proposed-label.south west) {};

  \node[
    anchor=north west,
    inner sep=0pt,
    font=\bfseries\tiny,
    text=luwarn
  ] at ([xshift=5pt,yshift=-4pt]proposal.north west)
    {STYLISTIC GOAL};

  \node[
    anchor=south west,
    inner sep=0pt,
    font=\bfseries\scriptsize
  ] at ([xshift=5pt,yshift=4pt]proposal.south west)
    {Keep sentences short};

  \node[
    anchor=north east,
    draw=luframe,
    fill=white,
    rounded corners=5pt,
    line width=0.4pt,
    inner xsep=4pt,
    inner ysep=1pt,
    font=\tiny,
    text=luink!70
  ] at ([xshift=-5pt,yshift=-4pt]proposal.north east)
    {New};

  \node[
    anchor=south east,
    fill=luaccent,
    text=white,
    rounded corners=3pt,
    inner xsep=6pt,
    inner ysep=3pt,
    font=\bfseries\tiny
  ] (add-button) at
    ([xshift=-7pt,yshift=14pt]f.south east)
    {Add as explicit intent anyway};

  \node[
    anchor=east,
    draw=luframe,
    fill=white,
    text=luink,
    rounded corners=3pt,
    line width=0.4pt,
    inner xsep=6pt,
    inner ysep=3pt,
    font=\tiny
  ] at ([xshift=-4pt]add-button.west)
    {Cancel};

  \node[
    anchor=south,
    font=\tiny,
    text=luink!55
  ] at ([yshift=3pt]f.south)
    {You stay in control --- the system only proposes; you decide.};

  \lubadge{f}{B}
\end{tikzpicture}} &
\begin{tikzpicture}[baseline=(f.north)]
  \node[lupanel,text width=\dimexpr\luD-8pt\relax,minimum height=\luH,inner sep=0pt] (f) {\phantom{x}};
  \node[lucard=\dimexpr\luD-22pt\relax,luderived,anchor=north] (c1) at ([yshift=-15pt]f.north) {\lutype{pragmatic effect}\\readers grasp the argument quickly\\\luweak};
  \node[lucard=\dimexpr\luD-22pt\relax,luexplicit,anchor=north] (c2) at ([yshift=-5pt]c1.south) {\lutype{stylistic goal}\\keep sentences short\\\luweak};
  \node[lucard=\dimexpr\luD-22pt\relax,luderived,anchor=north] (c3) at ([yshift=-5pt]c2.south) {\lutype{linguistic realization}\\one claim per sentence\\\luweak};
  \draw[-{Stealth[length=2.5pt,width=2.2pt]},luweakly,line width=0.5pt,dash pattern=on 1.5pt off 0.9pt on 0.4pt off 0.9pt] (c2.north) -- (c1.south);
  \draw[-{Stealth[length=2.5pt,width=2.2pt]},luweakly,line width=0.5pt] (c3.north) -- (c2.south);
  \lubadge{f}{C}
\end{tikzpicture} &
\begin{tikzpicture}[baseline=(f.north)]
  \node[lupanel,text width=\dimexpr\luE-8pt\relax,minimum height=\luH] (f) {%
    \textbf{Intent change saved}\\[1pt]
    \lutype{3 intents added}\\[-1pt]keep sentences short; one claim per sentence; readers grasp the argument quickly\\[2pt]
    \lutype{1 intent grounded}\\[-1pt]one claim per sentence, weakly satisfied\\[2pt]
    \lutype{3 fulfillment changes}\\[-1pt]not evaluated $\rightarrow$ weakly satisfied};
  \lubadge{f}{D}
\end{tikzpicture} \\
\end{tabular}
\par\vspace{2pt}
\begin{tikzpicture}[
  box/.style={fill=luweak,rounded corners=2pt,minimum height=12pt,inner xsep=5pt,inner ysep=1pt,font=\bfseries,text=luink,anchor=west},
  arr/.style={-{Stealth[length=3pt,width=2.6pt]},draw=luaccent,line width=0.5pt}]
  \useasboundingbox (0,0) rectangle (\textwidth,13pt);
  \node[box] (b1) at (\tabcolsep,6.5pt) {Action};
  \node[box] (b2) at (\dimexpr\luXb+\tabcolsep\relax,6.5pt) {Target};
  \node[box] (b3) at (\dimexpr\luXc+\tabcolsep\relax,6.5pt) {Confirmation};
  \node[box] (b4) at (\dimexpr\luXd+\tabcolsep\relax,6.5pt) {Local update};
  \node[box] (b5) at (\dimexpr\luXe+\tabcolsep\relax,6.5pt) {Impact summary};
  \foreach \a/\b in {b1/b2,b2/b3,b3/b4,b4/b5} \draw[arr] ([xshift=3pt]\a.east) -- ([xshift=-3pt]\b.west);
\end{tikzpicture}
\par\vspace{1pt}
\begin{tabular}{L{\luA} L{\luB} L{\luC} L{\luD} L{\luE}}
\midrule
\textbf{Unpin} & an explicit intent & preview & becomes derived & status changed \\
\midrule
\multirow[t]{3}{*}{\textbf{Pin}} & a derived intent, evaluated & preview & becomes explicit & status changed \\
\arrayrulecolor{lurule}\cmidrule[0.3pt]{2-5}
 & a derived intent, not yet evaluated & preview & becomes explicit; derive downward; propagate fulfillment; check conflicts & status changed; its fulfillment; intents appeared; conflict if any \\
\cmidrule[0.3pt]{2-5}
 & a hint & preview & becomes explicit, to be realized at the next regeneration & status changed \\
\arrayrulecolor{black}\midrule
\multirow[t]{3}{*}{\textbf{Edit}} & an explicit intent & preview & re-judge adjacent links; re-propagate; check conflicts & links changed type; fulfillment moved; conflict if any \\
\arrayrulecolor{lurule}\cmidrule[0.3pt]{2-5}
 & a derived intent, evaluated & preview & becomes explicit; re-judge adjacent links; re-propagate; check conflicts & status changed; links changed type; fulfillment moved; conflict if any \\
\cmidrule[0.3pt]{2-5}
 & a derived intent, not yet evaluated & preview & becomes explicit; re-judge adjacent links; derive downward; propagate; check conflicts & status changed; its fulfillment; intents appeared; links changed type; conflict if any \\
\arrayrulecolor{black}\midrule
\multirow[t]{4}{*}{\textbf{Add}} & below an existing intent & preview & new explicit intent; derive downward; re-propagate; check conflicts & intents appeared; fulfillment moved; conflict if any \\
\arrayrulecolor{lurule}\cmidrule[0.3pt]{2-5}
 & above an existing intent & preview & new explicit intent; derive upward; re-propagate & intents appeared; fulfillment moved \\
\cmidrule[0.3pt]{2-5}
\rowcolor{lurow}\lumarker & unrelated to any existing intent & \textcolor{luwarn}{warning: no relation} & new explicit intent; derive upward and downward; propagate fulfillment & its fulfillment; intents appeared \\
\cmidrule[0.3pt]{2-5}
 & working against an existing intent & \textcolor{luwarn}{warning: negative relation} & new explicit intent on a negative link; derive upward and downward; re-propagate; check conflicts & its fulfillment; intents appeared; fulfillment moved; conflict if any \\
\arrayrulecolor{black}\bottomrule
\end{tabular}
\caption{How an action updates the intent graph. Top: one example action's lifecycle. A comment is translated into adding a stylistic goal (A), which triggers a warning because it relates to no existing intent (B); on commit, the system derives intents upward and downward and propagates fulfillment (C) and reports the changes in an impact summary (D). Bottom: for each action on each target, the confirmation shown, the local update performed, and what the summary reports; the shaded row is the case illustrated above.}
  \Description{A figure in two parts. The top row follows one action through its lifecycle in four panels. A, a comment reading sentences too wordy right now is translated into adding keep sentences short, a stylistic goal. B, a confirmation dialog headed potential intent misalignment warns that the new intent has no obvious relationship to the writer's current intents and may shift attention away, shows the proposed new intent, and offers cancel or add as explicit intent anyway. C, the local update, in which the system derives a pragmatic effect above the new goal, readers grasp the argument quickly, and a linguistic realization below it, one claim per sentence, each with its own fulfillment value. D, an impact summary reporting three intents added, one intent grounded, and one fulfillment change from not evaluated to weakly satisfied. The bottom part is a table whose columns are action, target, confirmation, local update and impact summary. Rows cover unpin; pin, on a derived intent already evaluated, a derived intent not yet evaluated, or a hint; edit, on an explicit intent, a derived intent already evaluated, or a derived intent not yet evaluated; and add, below an existing intent, above an existing intent, unrelated to any existing intent, or working against an existing intent. The last two add rows carry warnings, no relation and negative relation. The shaded row is the case illustrated above.}
\label{fig:local-update}
\end{figure*}

\subsubsection{Conflict Resolution (DR2)} 
Each update also checks for conflicts, cases where two explicit intents are joined by a path along which advancing one works against the other. 
Both are commitments the writer has made, yet no draft can fully realize both. 
IntentWeaver therefore disables regeneration until the conflict is resolved, so that it cannot pass silently into the next draft. A banner explains the situation and offers three ways out (Figure~\ref{fig:storyboard} (3)). 
The system proposes editing one intent to a new value that eases the tension. 
Alternatively, the writer can unpin one side to let it lapse, or edit either by hand. 
Which intent to yield on, and how far, is a judgment IntentWeaver surfaces but does not make; it depends on which one the writer holds more firmly.

\subsubsection{Hints beyond the Draft (DR3)}
Everything in the graph so far is bound to the current draft. 
IntentWeaver additionally volunteers hints, intents that lie outside the draft but would advance an explicit intent, shown as cards tethered to it. 
A hint may attach to an intent the draft delivers weakly, suggesting what could fulfill it, or to one already \lvl{satisfied}, proposing a direction the intent could grow in beyond what the draft has settled for. 
Either way, it invites the writer to extend and evolve their intents beyond the current draft. 
A hint can be pinned, whereupon it becomes an explicit intent to be realized at the next regeneration, or dismissed; the writer can also refresh a hint for another, or request one for a particular intent.

\subsubsection{Regeneration and Final Editing (DR1)}  
After these actions, the writer can regenerate: IntentWeaver generates a new draft from the updated explicit intents, then derives a new intent graph from that draft and those anchors; derived intents that no longer describe the draft disappear and new ones take their place.
Each round begins from a firmer set of intents and a draft written to them. 
Version history keeps every round, traces how the explicit intents evolved from one to the next, and lets the writer restore any earlier version (Figure~\ref{fig:storyboard} (4)).
When the intents are clear and coherent and the draft close enough, the writer enters final editing and polishes the text by hand.

\providecommand{\system}{IntentWeaver}

\section{Evaluation}
To evaluate the effects of IntentWeaver, we conducted a mixed-methods, within-subjects study where participants completed co-writing tasks with the system and with a baseline AI writing tool. 
We explore \textbf{RQ1:} \emph{Does \system{} mitigate intent incoherence and support the evolution of high-level intents, and do these effects carry into the final text?}
We then turn to the writers' side, asking \textbf{RQ2:} \emph{How do writers appropriate \system{} in their co-writing process?} and
\textbf{RQ3:} \emph{How do writers experience co-writing with \system{}, in
terms of workload, usability, and creativity support?}

\subsection{Conditions and Tasks}
We developed a baseline modeled on commercial AI writing tools such as Lex \cite{lex2024} and Google Docs with Gemini \cite{google2024gemini}, retaining only three functions: generating a draft from the writer's initial prompt, requesting AI edits on a highlighted span or the whole draft, and editing the text directly. 
It adopts the same visual layout, styling, and underlying language model as \system{}, to minimize confounding variables.

Both conditions used social media writing tasks, a representative
real-world use case of human-AI co-writing \cite{reza2025co, lee2024design} widely adopted in
prior research \cite{kim2026intentflow,reza2024abscribe}. We crafted the two tasks around three
requirements. First, they required no domain expertise, so that performance
would reflect participants' writing process. Second, each task specified a concrete communicative situation, including the persona, the triggering event, and the audience, to anchor participants in a shared context. Third, both tasks were open-ended: the
briefs deliberately withheld the purpose and stance, so that participants had
to form their own intents during writing and could revise them as the draft
took shape.
In one task, the participant writes a LinkedIn post after their company ends
the internship program they had mentored for; in the other, they post to a
running community they organize after a rule change drove some early members
away. 
Pilot testing further verified that the two tasks were
comparable in complexity and in the range of viable communicative purposes.
The full briefs are provided in the supplementary material.

\subsection{Participants and Procedure}

We determined the sample size through an a priori power analysis in G*Power
\cite{faul2007g}: for within-subjects comparisons at $\alpha=.05$, detecting a
large effect ($d=0.8$) with $90\%$ power requires at least 20 participants.
With IRB approval, we recruited 24 participants via social media, online
advertisements, and word-of-mouth (10 female, 14 male, aged 19--32; backgrounds
spanning engineering, science, arts, and business). Eligibility required
frequent experience of co-writing with AI, and we screened out those who had
taken part in the formative study to avoid bias.

\begin{figure*}[t]
  \centering
  \includegraphics[width=0.93\textwidth]{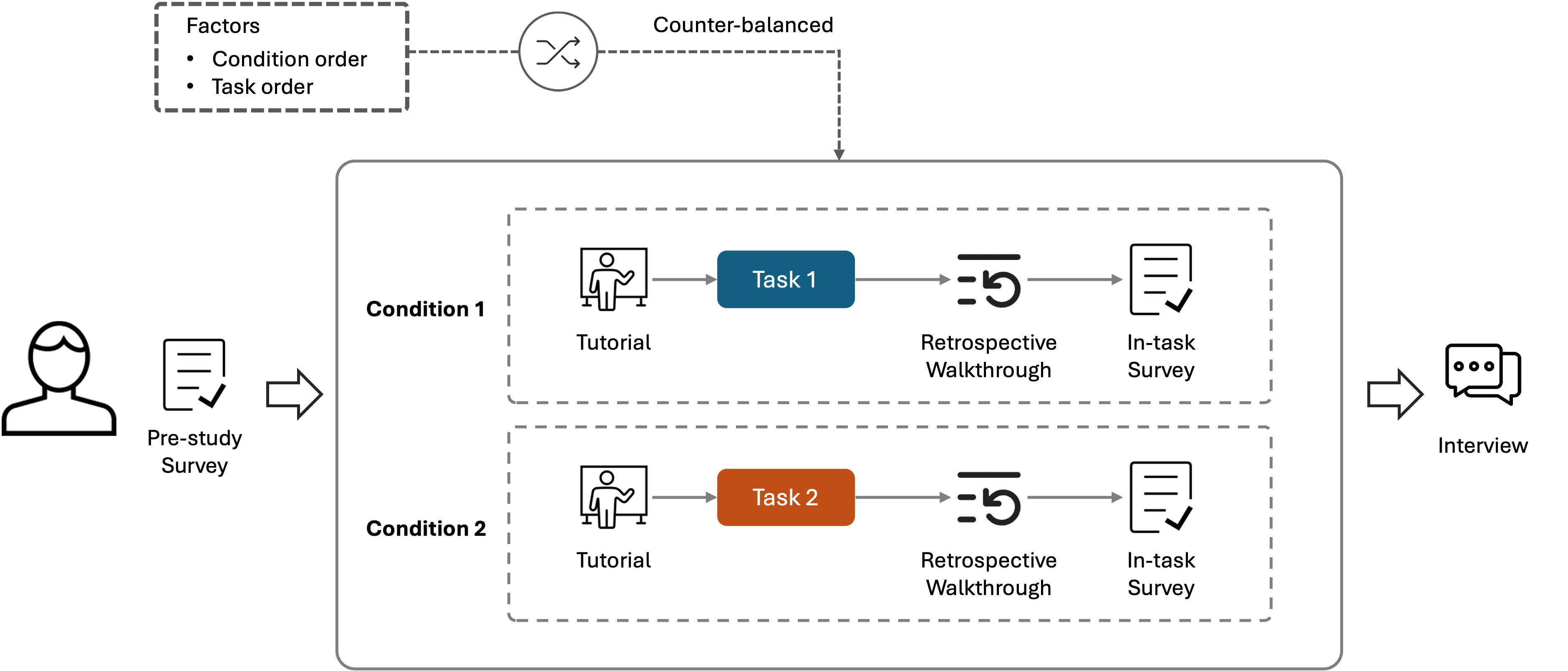}
  \caption{Study procedure. After a pre-study survey and briefing, participants completed two co-writing tasks in a fully counterbalanced order, one with \system{} and one with the baseline. Each task was preceded by a tutorial and followed by a retrospective walkthrough and a questionnaire; the session closed with a semi-structured interview.}
  \Description{A flowchart of the study procedure. A participant icon leads into a pre-study survey and then into a box holding two conditions. Condition one runs a tutorial, then task one, then a retrospective walkthrough, then an in-task survey. Condition two repeats the same four steps with task two. A note above the box records that condition order and task order were counter-balanced. An arrow out of the box leads to a closing interview.}
  \label{fig:procedure}
\end{figure*}

Figure~\ref{fig:procedure} shows the study procedure. After a demographic survey and a briefing, participants completed the two tasks in a fully counterbalanced design, one with \system{} and one with the baseline, each preceded by a tutorial in which they could explore the system and ask questions. They were given 25 minutes per task and asked to articulate their writing intent when they finished. Each task was followed by a retrospective walkthrough \cite{van2005uncovering}, in which participants revisited each action and explained their thinking at the time, and a questionnaire. Sessions ended with a semi-structured interview about how participants worked with \system{}, why they used particular features, and how the two systems compared.
Screen activity, audio, and system logs were captured with
participants' consent. A session took around 120 minutes, and compensation
followed the local hourly wage.

\subsection{Measurements}
\label{sec:measures}

We evaluated each RQ using behavioral measures from session data, expert
ratings, and self-reported measures. All questionnaire items and expert ratings used a 7-point scale (except intent fulfillment using a 5-point scale). The specific items and full annotation pipeline are provided in the supplementary material.

\subsubsection{RQ1: Does \system{} mitigate intent incoherence and support
the evolution of high-level intents, and do these effects carry into the final
text?}
To answer RQ1, we observed effects from three angles: participants'
behavior and cognition during writing, the intents they articulated, and the
text they produced.

During writing, we measured two behavioral indicators. The \emph{drift
detection rate} is the probability that an action drifting from the
writer's high-level intents was noticed when it occurred. In the retrospective
walkthrough, participants judged for each action whether it had served a
high-level intent they held at that moment, and, for actions that had not,
whether they had noticed this at the time; the rate is the share of these
drifting actions that participants reported noticing. 
The \emph{pop-back-up rate} is the number of transitions from an action serving low-level intents directly to one serving high-level intents, normalized by time on task. To compute it, we first extracted a stream of actions that changed the text or the intents behind it (excluding mechanical steps such as restore and export) from the interaction logs of both conditions. Two researchers independently coded the intent level each action served, based on the definitions in Section~\ref{sec:hierarchy} (\textsc{low} for local content purpose, content unit, stylistic goal, and linguistic realization; \textsc{high} for all other types), and resolved discrepancies through
discussion. Four questionnaire items complemented these measures with
participants' subjective perception of their own writing behavior.

\begin{figure*}[t]
  \centering
  \includegraphics[width=\textwidth]{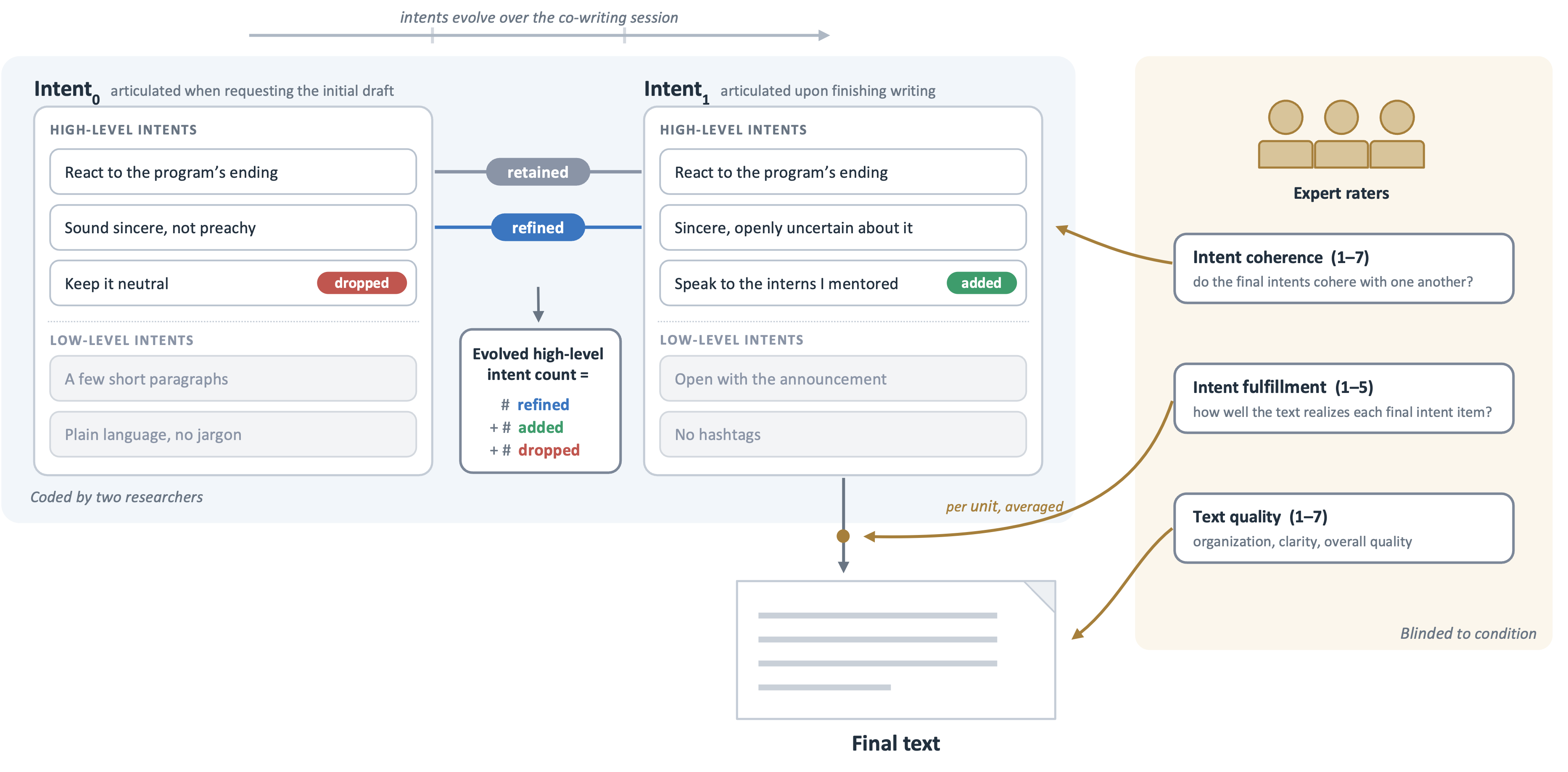}
  \caption{Coding of intent evolution and expert ratings of the final outcome. Two researchers coded Intent\textsubscript{0} (when requesting the initial draft) and Intent\textsubscript{1} (upon finishing writing); high-level units were labeled \lvl{retained}, \lvl{refined}, \lvl{added}, or \lvl{dropped}, and the evolved high-level intent count is their sum excluding \lvl{retained}. Blinded expert raters then scored the internal coherence of Intent\textsubscript{1}, how well the final text fulfills each remaining intent, and the text's organization, clarity, and overall quality.}
  \Description{A diagram of the RQ1 coding and rating pipeline, with a timeline across the top reading intents evolve over the co-writing session. On the left, Intent zero, articulated when requesting the initial draft, lists high-level intents such as react to the program's ending, sound sincere not preachy, and keep it neutral, together with low-level intents such as a few short paragraphs and plain language, no jargon. On the right, Intent one, articulated upon finishing writing, lists react to the program's ending, sincere and openly uncertain about it, and speak to the interns I mentored, with its own low-level intents. Labels between the two mark units as retained, refined, added or dropped, and a box defines the evolved high-level intent count as the number refined plus added plus dropped. Both articulations were coded by two researchers. An arrow leads from Intent one down to the final text. A panel on the right shows three expert raters, blinded to condition, supplying three measures: intent coherence on a one to seven scale, whether the final intents cohere with one another; intent fulfillment on a one to five scale, how well the text realizes each final intent unit, averaged per unit; and text quality on a one to seven scale, covering organization, clarity and overall quality.}
  \label{fig:metric-explanation}
\end{figure*}

In the articulated intents, we measured the \emph{evolved high-level intent
count} (Figure~\ref{fig:metric-explanation}). For each task, we compared the participant's articulation of intent when
requesting the initial draft (Intent\textsubscript{0}) with their articulation
upon finishing writing (Intent\textsubscript{1}). Both were segmented into
intent units and coded for level by two researchers, following the same
procedure as above. Units expressing the same intent in both articulations
were then paired, while units present in only one remained as single-sided
entries; each pair or entry was coded as \lvl{retained}, \lvl{refined},
\lvl{added}, or \lvl{dropped}. The count is the number of high-level entries
coded as anything other than \lvl{retained}, capturing all movement in a
participant's high-level intents over the session. We additionally measured
the \emph{internal coherence} of Intent\textsubscript{1}, whether the intents
a participant ended with still cohere with one another, rated by the expert
panel described next.

For the final text, we recruited three professional writers with experience
writing social media content for real audiences (one with five years, two
with ten). Blinded to condition, they received each text together with its
coded Intent\textsubscript{1} unit set and independently rated three things:
the internal coherence of the intent set; how well the text realizes each
intent unit, with scores averaged across raters and units to yield a
per-text \emph{intent fulfillment} score, reported overall and separately for high-/low-level units; and the text's \emph{organization}, \emph{clarity}, and \emph{overall quality}.

\subsubsection{RQ2: How do writers appropriate \system{} in their co-writing
process?}
To understand how writers actually worked with \system{}, we analyzed the
complete interaction logs of all \system{} sessions, identifying which
features participants used and the usage patterns they formed. Participants also rated the usefulness of each feature in the questionnaire, and the semi-structured interviews probed the rationale behind their usage.

\subsubsection{RQ3: How do writers experience co-writing with \system{}, in
terms of workload, usability, and creativity support?}
We measured task workload on four items adapted from the NASA Task Load Index
\cite{hart1988development}, covering \emph{mental demand}, \emph{temporal demand},
\emph{effort}, and \emph{frustration}. Usability on four items adapted
from the System Usability Scale \cite{brooke1996sus}, covering \emph{ease of
learning}, \emph{ease of use}, \emph{integration of functions}, and
\emph{willingness for future use}. We further assessed \emph{exploration} and
\emph{worth the effort} from the Creativity Support Index \cite{cherry2014quantifying},
together with participants' \emph{sense of agency} \cite{limerick2014experience}.

\begin{table*}[t]
  \caption{Overview of measures, grouped by the research question they serve.
  Full questionnaire items, the annotation pipeline, and participant and expert rater backgrounds are provided in the supplementary material.}
  \label{tab:measures}
  \Description{A three-column table listing every measure, its source, and its description or questionnaire item, grouped into three bands by research question. The RQ1 band lists the drift detection rate and the pop-back-up rate from behavioral data, four questionnaire items covering detail fixation, local absorption, unclear purpose and goal propagation, the evolved high-level intent count from the interaction logs, and three expert ratings covering intent coherence, intent fulfillment and text quality. The RQ2 band lists usage patterns from the interaction logs and feature usefulness from the questionnaire. The RQ3 band lists task workload, usability and creativity support, each a set of questionnaire items.}
  \centering
  \includegraphics{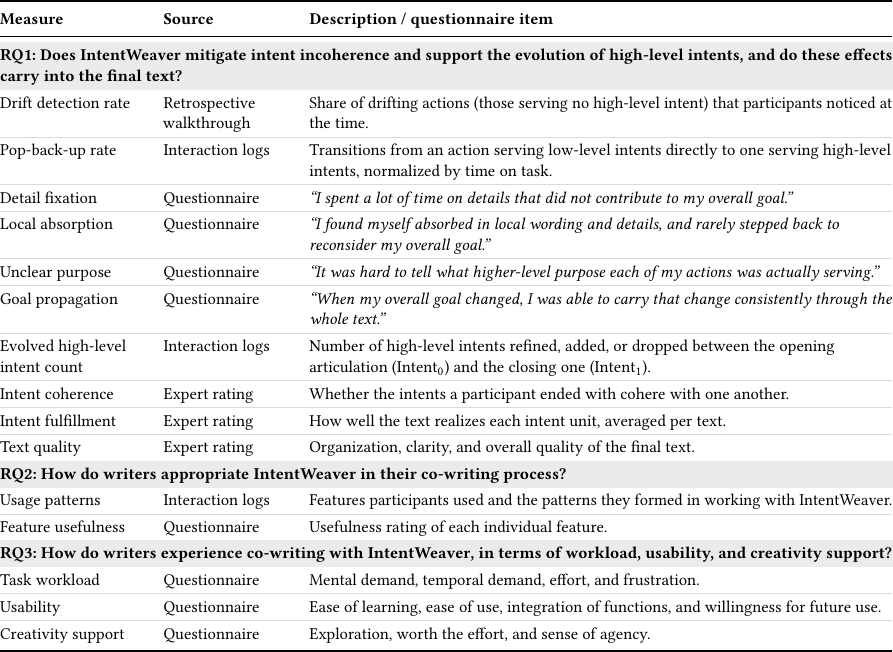}
\end{table*}

\subsection{Analysis Methods}
\label{sec:analysis}

We analyzed the data with a mixed-methods approach. For quantitative measures, Shapiro--Wilk tests \cite{shapiro1965analysis} showed that most variables departed from normality, so all within-participant comparisons used the two-sided Wilcoxon signed-rank test \cite{woolson2007wilcoxon},
for which we report the test statistic, $p$-value, and effect size $r$ \cite{sullivan2012using,tomczak2014need,rosenthal1994parametric}. 
This applies to all measures of RQ1 and RQ3 except the drift detection rate, which is undefined for participants who made no drifting action. We therefore computed it over all drifting actions in each condition and compared the two conditions with a permutation test \cite{ernst2004permutation} that derives the null distribution by reshuffling condition labels within each participant, preserving the paired structure of the design.
For expert ratings, we also report inter-rater reliability with the
intraclass correlation coefficient ICC(2,3) \cite{koo2016guideline,shrout1979intraclass}, which treats the three raters as a random sample of possible raters and estimates the reliability of their averaged score.
Descriptive statistics summarize the usage logs and feature ratings for RQ2.

For qualitative data, we performed a thematic analysis
\cite{braun2006using} of the interview transcripts. Two researchers coded them independently in ATLAS.ti \cite{friese2014qualitative}, organized codes by the RQ they addressed, and met regularly to
reconcile discrepancies until reaching consensus. 
The resulting themes were refined through discussion among all authors and are reported with
representative quotes alongside the quantitative results in
Section~\ref{sec: results}.

\providecommand{\system}{IntentWeaver}

\section{Results}
\label{sec: results}

This section reports the findings, moving from the effects of \system{} on intent and text (RQ1), to how participants appropriated (RQ2) and experienced it (RQ3). Statistical results anchor each finding, enriched by interview accounts.

\subsection{RQ1: Mitigating Intent Incoherence and Supporting Intent Evolution}
\label{sec:rq1}

We first examine the co-writing process, and whether the effects observed there are
reflected in the outcome. Detailed statistics are presented in Table~\ref{tab:rq1}.

\providecommand{\system}{IntentWeaver}
\providecommand{\dn}{\,$\downarrow$}
\providecommand{\up}{\,$\uparrow$}
\providecommand{\pmark}[1]{\textsuperscript{#1}}
\providecommand{\best}[1]{\textbf{#1}}

\begin{table}[t]
  \centering
  \caption{\textbf{Statistical analysis of RQ1 comparing the baseline and \system{}}. The drift detection rate is pooled over drifting actions and compared with a within-participant permutation test, with the rate difference as effect size. All items use a 7-point scale, except intent fulfillment using a 5-point scale; the pop-back-up rate is transitions per ten minutes on task. Inter-rater reliability ranged from ICC(2,3) $=.59$ (overall quality) to $.87$ (intent coherence). \up{}/\dn{} mark whether higher or lower is better; the better condition is \textbf{bold} where $p<.05$ ($-$: $p>.10$; $+$: $.05<p<.10$; \pmark{*}$p<.05$; \pmark{**}$p<.01$; \pmark{***}$p<.001$).}
  \Description{A table comparing the baseline and IntentWeaver on four blocks of RQ1 measures, reporting for each item the mean and standard deviation per condition, a Wilcoxon Z statistic, a p value and an effect size. The behavioral process block holds the drift detection rate and the pop-back-up rate. The self-report process block holds detail fixation, local absorption, unclear purpose and goal propagation. The articulated intents block holds the evolved high-level intent count and the expert-rated coherence of the articulated intent set. The final text block holds overall intent fulfillment, high-level and low-level fulfillment, organization, clarity and overall quality. IntentWeaver is significantly better on drift detection, pop-back-up rate, local absorption, unclear purpose, goal propagation, evolved high-level intent count, overall and high-level intent fulfillment, and clarity; organization, low-level fulfillment and overall quality do not differ.}
  \label{tab:rq1}
  \includegraphics{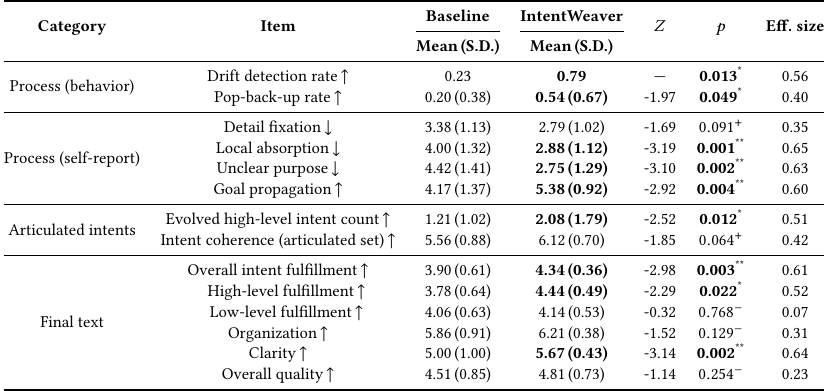}
\end{table}

\subsubsection{During the process}
During co-writing, participants with \system{} more readily recovered from the dynamics that
undermine intent coherence. They \textbf{popped back up} from local work to their high-level
intents at a significantly higher rate and reported being markedly less \textbf{absorbed in local
wording}. The interviews trace this first to the hierarchy staying in view. In the baseline,
participants often lost hold of their opening prompt once the draft appeared and revised
\textit{``without checking against my original intent''} (P03); intents that are not
externalized, P06 observed, \textit{``don't stay in working memory.''} With the hierarchy
beside the text, overarching intents sit at the top of the graph where the eye lands
first (P10, P13), and several participants adopted a top-down routine of checking whether each
intent was fulfilled before reading the text (P03, P04, P06). The read-only draft reinforced this: unable to edit directly, participants found
themselves \textit{``forced to prompt at high level''} (P04) instead of \textit{``picking
at words''} (P05). 

When an action made the hierarchy incoherent, participants with \system{} were
significantly more likely to \textbf{detect the drift} at the time, and they reported being
much \textbf{clearer about what each action meant for their higher-level intents}. Eight participants attributed this clarity to the chain the system makes visible from a passage up to the high-level intents it serves, along which they could see what any operation would serve, leave untouched, or work against.
The pre-action confirmation and impact summary surfaced this chain at the moment of
acting (P10, P12). The confirmation made P10 \textit{``think about whether I really was
misaligned,''} and even when P02 found its warning imprecise, it still made him \textit{``think
one layer further.''}

Reconciliation also ran in the top-down direction. After revising a high-level intent,
participants reported being reliably better able to \textbf{propagate the change consistently}
through the rest of the hierarchy and the text. Changing an intent at the top automatically
adjusted all related content at once, leaving participants only to check the result (P08,
P23), whereas the baseline required \textit{``spelling out everything"} (P13).

\subsubsection{Final outcome}
These effects are reflected in what participants ended with. Over a session, participants
evolved significantly more of their \textbf{high-level intents} with \system{}, and they
located this in two elements of the hierarchy. The derived intents articulated participants' tacit thoughts and what the draft had introduced, helping them tell which they wanted and which they did not (P05, P06). The links prompted them to
sort out and reflect on their goals actively (P02, P05), in both directions: a clarified
low-level intent \textit{``propagates clearly up to the high level''} (P04), while
seeing what a change would do to lower levels \textit{``grounds a free-floating idea in concrete effects,''} so that P12 evolved his intents \textit{``more rationally.''} 
As intents kept evolving, the hierarchy stayed coherent. Expert-rated \textbf{coherence of the intent sets} was marginally higher with \system{} with a medium effect, and the texts bore out the same pattern more reliably in concrete terms: baseline texts fulfilled low-level intents better than high-level ones (4.06 vs.\ 3.78), whereas with \system{} \textbf{high-level fulfillment} rose to 4.44 while \textbf{low-level fulfillment} held steady (4.06 to 4.14), so the high-level intents, which lagged in baseline texts, were now realized at least as well as the local ones.
Overall, the final texts also \textbf{fulfilled participants' intents} significantly
better. Among the quality ratings, \textbf{organization} and \textbf{overall quality} did
not differ, but \textbf{clarity} was significantly higher with \system{}. Clarity is sensitive to whether a text commits to a definite purpose, and P18's praise for the output pointed the same way: \textit{``the logic is clearer, I have a better grip on the theme.''}

\begin{figure*}[t]
  \centering
  \includegraphics[width=\textwidth]{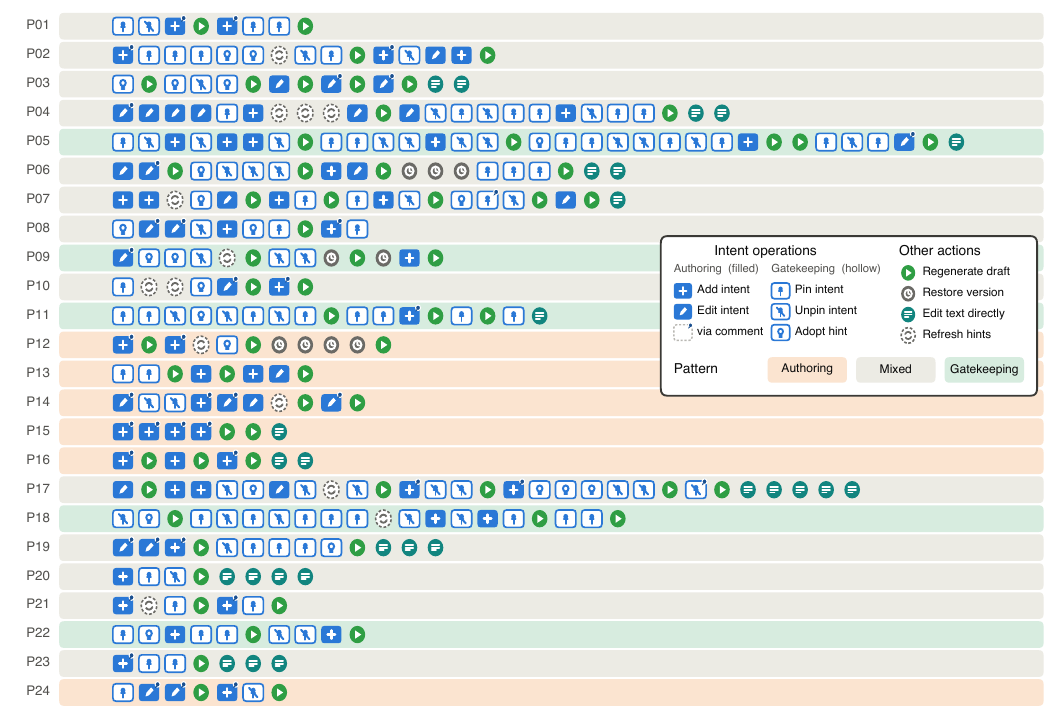}
  \caption{All 24 \system{} sessions as ordered action sequences, one row per participant, colored by usage pattern. Hollow chips are gatekeeping actions (pin, unpin, adopt hint); a speech-bubble badge marks an operation translated from a comment, the rest were performed on the graph directly. 
    }
  \Description{Twenty-four rows of small chips, one row per participant from P01 to P24, showing every action in the order it occurred. Square chips are operations on the intent hierarchy: filled squares are authoring actions, adding or editing an intent, and hollow squares are gatekeeping actions, pinning, unpinning or adopting a hinted intent. A speech-bubble badge marks an operation that arrived by translating a comment rather than acting on the graph directly. Round chips are the other actions: regenerating the draft, restoring a version, editing the text directly, and refreshing hints. Row backgrounds are tinted by usage pattern, six rows authoring, thirteen mixed and five gatekeeping, assigned by the share of gatekeeping actions among a participant's intent operations. Direct text edits, where present, form the final block of a row. A three-column legend sits inside the figure, over the shorter middle rows.}
  \label{fig:rq2-usage}
\end{figure*}

\subsection{RQ2: How Writers Appropriated \system{}}
\label{sec:rq2}

Beyond whether \system{} worked, we asked how participants worked with it. The interaction
logs show where each participant put their effort between writing intents, ruling on the
system's inferences, and editing the text (Figure~\ref{fig:rq2-usage}); the ratings show which
features they credited (Figure~\ref{fig:rq2-diverging}); and the interviews explain both.

\subsubsection{Usage patterns}
\label{sec:rq2-usage}
 
Operations on the intent hierarchy fall into two kinds. In an \emph{authoring} action (add,
edit) the writer states an intent in their own words; in a \emph{gatekeeping} action (pin,
unpin, adopt hint) the writer commits to or releases an intent the system has worded. Actions also
travelled through different channels, half of the authoring actions arriving by translating a
comment and almost all gatekeeping actions being performed on the graph directly. The actions
reflect different ways of treating the hierarchy, as a reference against which to state what one
wants or as a pool of candidates to select from, which led us to distinguish three usage
patterns by the share of them among a participant's intent operations, cut at natural gaps in its distribution (Figure~\ref{fig:rq2-usage}).
 
\patA{\textbf{Authoring}} participants (6/24) wrote most of their intents themselves and made the fewest
intent operations (median 4.5). 
For them the work happened on their own intents and the text; the graph only recorded it.
They worked outward from explicit intents they already held, supplying what the draft still lacked (P13, P15, P24) and paying little attention to derived intents. Two reasons recur. Reading the derived layer
was effortful. And ignoring it cost nothing, because a derived intent unadopted \textit{``doesn't affect the text''} (P24). Even so, derived intents served this group as a menu to explore: P15 saw \textit{``more directions to revise''} than in the text alone, and P12 found it added \textit{``things I had never thought of,''} which he then wrote up in his own terms.
 
\patG{\textbf{Gatekeeping}} participants (5/24) made the most intent operations (median 13) and spent them mainly ruling on what the system had derived. They share one attitude: the hierarchy was the object to work on while the text was just its rendering. P11 reasoned that if \textit{``each node's purpose matched mine''} the sentences would follow. This attitude may account for two observations. The largest gain in pop-back-up rate between conditions sits in this group (+0.68 against +0.37 and $-$0.01 in the other two) since a writer who works on the hierarchy is naturally drawn toward its top, where the graph places overarching intents. And no gatekeeping participant voiced the wish to edit the draft directly that many others expressed; the wording had been delegated.
 
\patM{\textbf{Mixed}} participants (13/24) did both in comparable measure (median 8 operations). Their accounts describe the two kinds of action as halves of one inspection routine: after
a generation they went through the derived intents first, committing to those that stated
\textit{``what I meant, more explicitly than I had''} (P01) and releasing those they did not
want (P19), and then wrote the change that the derived layer had not supplied.
 
Besides the intent operations, we examined whether participants, on entering final
editing mode, edited the text or submitted it directly. The sample split in half. Twelve
finished without a single edit, trusting the hierarchy to carry the wording: \textit{``if the
cards are set right, I trust it to generate along that logic''} (P21). The other twelve kept the last word for themselves, noting during exploration the phrasings they could not yet touch, and fixing them there.

\subsubsection{Perceived usefulness of features}
\label{sec:rq2-features}

\begin{figure}[t]
  \centering
  \includegraphics[width=\linewidth]{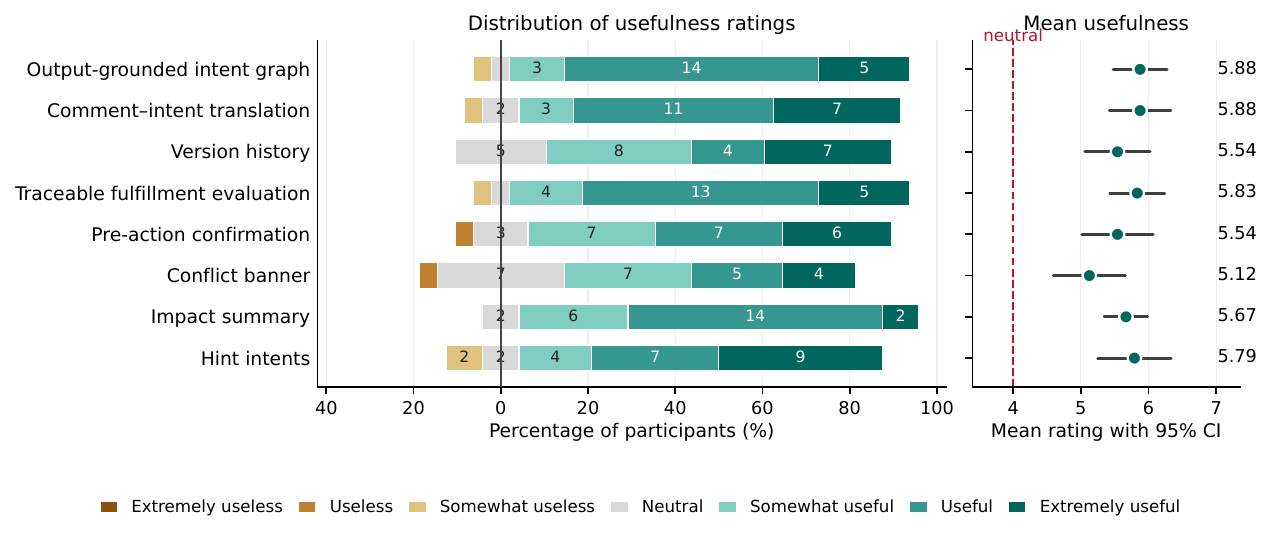}
  \caption{Perceived usefulness of each \system{} feature. Left: rating distribution diverging around the neutral midpoint, with participant counts; right: mean rating with 95\% CI, dashed line at neutral (4).}
  \Description{A two-panel chart of the usefulness ratings for the eight IntentWeaver features. In the left panel each feature has a horizontal bar showing its rating distribution diverging around the neutral midpoint, with segments colored from extremely useless through neutral to extremely useful and participant counts printed inside the larger segments. The features, top to bottom, are output-grounded intent graph, comment-intent translation, version history, traceable fulfillment evaluation, pre-action confirmation, conflict banner, impact summary, and hint intents. The right panel plots each feature's mean rating with a 95 percent confidence interval against a dashed line at the neutral midpoint of four. Every mean lies above neutral, from 5.12 for the conflict banner to 5.88 for the intent graph and for comment-intent translation.}
  \label{fig:rq2-diverging}
\end{figure}

All eight features were rated above the neutral midpoint (means 5.12--5.88,
Figure~\ref{fig:rq2-diverging}). The interviews show what each was used for and where it fell short.

The \textbf{output-grounded intent graph} received the highest rating, and many participants
(14/24) named it as what they valued most. It externalized their \textit{``line of reasoning"} (P09). Its split into rhetorical and content intents matched how participants conventionally think, so the
graph did the sorting they would otherwise do in their heads (P02, P08), or do poorly (P07);
P02, for instance, could \textit{``focus on one category of intent at a time.''} Two limits
recur. Some participants described the graph growing crowded across
iterations, and P14 proposed unfolding it level by level instead of all at once. Others found
manipulating the graph less free than editing text (P17, P22).
\textbf{Comment--intent translation} was rated as highly. A casual comment came back as a
well-formed intent at the right level, which made participants aware of tacit thoughts and kept them using the channel to evolve intent (P16, P20). However, P07 avoided the channel throughout for fear it \textit{``would
mess up my graph''}. 
\textbf{Version history} earned P07's only top score, because \textit{``after each update I could be lost and need a way back''}. P12 used it to trace which intent change had caused which text change.

\textbf{Traceable fulfillment evaluation}, the hover-and-highlight link between intents and
text, was praised by nine participants as an audit instrument: hovering sentence by sentence
to see \textit{``which of my goals it serves and whether all were covered''} (P03), or reading with one eye on which cards light up (P24). Seeing the correspondence made them more willing to trust the generation (P21, P24). P01, however, refused to delegate the evaluation: she would \textit{``judge whether it's satisfied myself, not have the model grade itself.''} \textbf{Pre-action confirmation} was described as the moment of reflection in Section~\ref{sec:rq1}; P17 additionally wanted to adjust the proposed change
inside the dialog instead of only confirming or cancelling. The \textbf{conflict banner} received the lowest rating and the most neutral ones, likely because many participants never triggered it. Those who did thought highly of it. P12, who called himself insensitive to such clashes, valued being told that \textit{``two points are fighting,''} and P13 used it to check \textit{``whether my thinking has wandered off topic.''}

The \textbf{impact summary} let P12 \textit{``know what I'm changing and whether it affects my expression.''} \textbf{Hint intents} were rated highly even though ten participants never adopted one; what participants valued
was being shown a dimension beyond the current text, whether or not they took it. P19 credited hints with making \system{} \textit{``better at divergent thinking,''} and P07 valued a dimension he \textit{``still didn't want but had to consider.''}

\begin{figure}[t]
  \centering
  \includegraphics[width=0.9\linewidth]{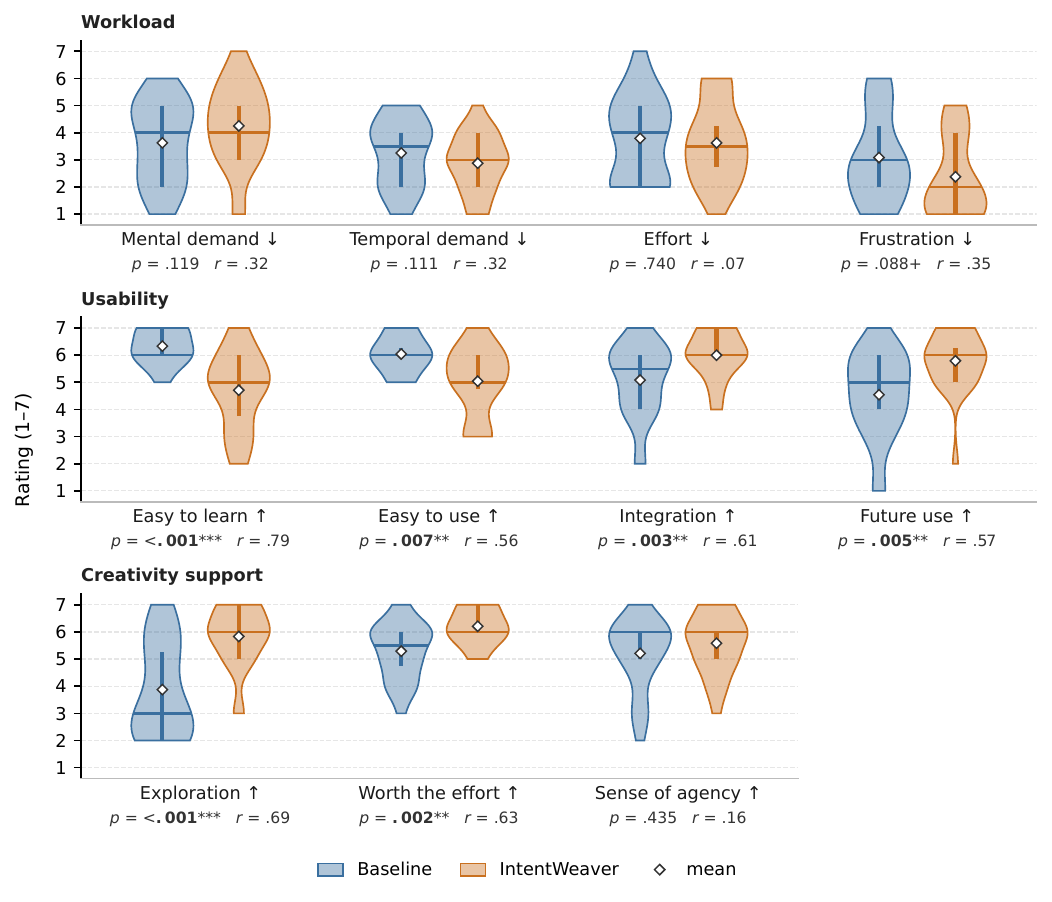}
  \caption{Task workload, usability, and creativity support by condition ($N=24$, 7-point
    items). Violins show the rating distribution, the horizontal line the median, and the diamond the mean. Below each item:
    $p$ of a two-sided Wilcoxon signed-rank test ($^{+}p<.10$; $^{*}p<.05$;
    $^{**}p<.01$; $^{***}p<.001$), in bold where $p<.05$, and effect size
    $r=|Z|/\sqrt{N}$. $\uparrow$/$\downarrow$: higher/lower is better.}
  \Description{Three panels of paired violin plots, one pair per questionnaire item, the left violin for the baseline and the right violin for IntentWeaver. Within each violin a horizontal line marks the median, a vertical bar the interquartile range, and a small diamond the mean. The workload panel holds mental demand, temporal demand, effort and frustration, where lower is better; the usability panel holds ease of learning, ease of use, integration of functions and willingness for future use, where higher is better; the creativity support panel holds exploration, worth the effort and sense of agency, where higher is better. Below each item are the p value of a two-sided Wilcoxon signed-rank test, set in bold where p is below .05, and the effect size r. No workload item differs. The baseline is rated higher on ease of learning and ease of use. IntentWeaver is rated higher on integration of functions, willingness for future use, exploration and worth the effort. Sense of agency does not differ.}
  \label{fig:rq3-boxes}
\end{figure}

\subsection{RQ3: How Writers Experienced \system{}}
\label{sec:rq3}

Figure~\ref{fig:rq3-boxes} reports task workload, usability, and creativity support.
No \textbf{task workload} item differed statistically between conditions, suggesting that even with
its added features, \system{} kept workload on par with the baseline. Some participants
nonetheless experienced the thinking the hierarchy calls for as extra load: \textit{``the baseline has no such step; this system makes me think about intent systematically''} (P02).

Regarding usability, \system{} was rated significantly harder to \textbf{learn} and to
\textbf{use}. The interviews paint a more differentiated picture. Nine participants agreed that
the system was hard to learn but fine to use once learned. What had to be learned was the loop
itself - that a change is made on the hierarchy, checked, and only then regenerated into text
(P19) - and the graph structure (P12, P17). Once learned, the flow felt familiar, since revising intents in response to output is what they already do when co-writing with AI (P14, P16, P20). Moreover, participants rated the functions significantly better \textbf{integrated} and were significantly more \textbf{willing to use it in the future}. As P03 put it, \textit{``from top intents down to sentences, from confirmation to impact summary and conflict banner, everything is linked.''} The learning cost did not deter them: P02, P23 and P24 rated
learning much harder and still rated future use four to five points higher.

On creativity support, participants rated \system{} significantly higher on \textbf{exploration}, which came from hints and derived intents showing \textit{``new directions I might add''} (P24), and judged the results more \textbf{worth the effort}. P02, who reported the extra step, also found it repaid: \textit{``an extra step, but the quality afterwards is higher.''}
\textbf{Sense of agency} did not differ, and the interviews show two effects cancelling. Some
participants felt a firmer hold on how the piece developed, \textit{``a stronger sense of
control over the whole process''} (P13). They noted that derived intents are inferred around their own explicit ones, so that exploring them extended their own direction step by step, never jumping away from it (P22, P24). Others felt control slip at the level of operation, mainly over the editing constraint: \textit{``it feels out of control; I still want to edit it myself''} (P04). Even so, they conceded that the constraint kept their attention on the higher levels (P04, P14).

\section{Discussion}

Our evaluation shows that \system{} helps writers evolve their intents while keeping them coherent. In this section we explore the possible reasons behind this outcome and its side effects, draw design considerations for similar systems, and discuss where the approach applies and where our study falls short.

\subsection{Intent as an Emergent Structure}
\label{sec:disc-emergent}

Prior work has transformed intent from natural-language prompts into structures that writers can inspect and manipulate~\cite{kim2026intentflow,siddiqui2025script,zhang2023visar}, helping them form and articulate what they want. These structures, however, are built before generation, and once the text exists, structure and text go their separate ways. This overlooks a basic property of intent: it is emergent. Intent is formed, negotiated, and attributed in the course of interaction, not fixed before it~\cite{stamp1990construct}; in writing specifically, goals grow through composing the text~\cite{flower1981cognitive}. 
However well an intent has been developed in advance, new lower-level intents keep emerging as the writer interacts with AI while higher-level intents stay frozen, drifting into incoherence without noticing~\cite{jakesch2023co, bhat2026reactive}.
PF6 put it plainly: \textit{``I spent a lot of effort writing a detailed prompt; the AI still led me off course.''} 
\system{} therefore reads intent from the text after each generation, making the text the medium through which the writer evolves their intent.

Our results indicate that with \system{} the mechanisms for maintaining intent coherence were restored, high-level intents evolved further, and the final text realized the writer's intents better.  
We interpret this from a cognitive angle. Bereiter and Scardamalia distinguish two ways of writing~\cite{bereiter1987psychology}. In knowledge telling, the writer retrieves what they know and sets it down as text in order; thinking runs one way, the load is low, and this is typical of novices. 
In knowledge transforming, the writer reworks and restructures their ideas while
writing, moving between what to say and how to say it to whom; the load is higher, and this is how skilled writers write. Co-writing with an LLM pushes writers toward telling. Generation arrives in bulk, and the writer's work becomes accepting content and patching its surface, which consumes the bandwidth that transforming would require~\cite{bhat2026reactive, buschek2021impact}; even professional writers report being carried along and losing their voice~\cite{ippolito2022creative, gero2023social}. 
\system{} pushes writers back toward transforming. First, it casts the draft as a medium for exploring intent, so that a writer is
drawn into reflecting on their goals in the very act of taking in the text. Second, the system relieves the writer of monitoring, externalizing and restructuring intent, which frees bandwidth for thinking. 
Third, the hierarchy itself is adopted from the transforming model, so the writer's dialogue on it follows the shape of that deeper thinking.
Where telling typically leaves a writer with no new insight, transforming deepens understanding, and this is what participants described: knowing more clearly what they wanted (P04, P07). 

The same logic has a side effect. We meant to keep writers from becoming trapped in the
details of the text, but some participants reported paying it almost no attention at all (P03--07, P20). 
The knowledge-constituting model of writing holds that some ideas arise only in concrete text and cannot be derived from a plan, since wording carries associations and particulars that no statement of its intent preserves~\cite{galbraith1999writing}. 
A writer who works only on the structure loses this route. 
Systems of this kind should therefore also draw the writer back to reading.
Passages left unread since the last generation, or sentences rewritten by a regeneration, can be marked for the writer's attention.

\subsection{Design Considerations}
\label{sec:disc-dc}

\subsubsection{Keep AI-inferred intent conservative and subordinate to the writer's own}
\label{sec:7.2.1}
A notable result is that participants explored significantly more with \system{}
while their sense of agency did not drop (Section~\ref{sec:rq3}), and they described the direction of their piece as firmly in hand, refined step by step without jumps (P22, P24). This contrasts with the observations
in Section~\ref{sec: 2.1} that conventional co-writing with AI sways writers' stance without their
noticing~\cite{jakesch2023co}. We attribute the difference to how \system{} infers intent. The
derived hierarchy is itself AI output, and it could have been produced exhaustively, surfacing
every dimension along which the text can be read, much as Luminate enumerates the dimensions
of a design space from a prompt~\cite{suh2024luminate}. \system{} instead expands the hierarchy only around the intents the writer has stated, renders inferred intents visually secondary, and lets
nothing derived drive the next generation unless the writer adopts it. The reason is that an
exhaustive layer would confront the writer with dimensions they never cared about, and the
attention capture that occurs at the level of sentences would recur at the level of intent as
choice overload~\cite{iyengar2000choice}. Writers would then evolve their intents toward what the AI
had proposed, and the intent hierarchy would become one more place where control slips away.

We therefore recommend that systems which infer intent from a writer's material, whether from generated text, from the writer's own drafts, or from their edits, keep the inferred layer conservative and clearly subordinate to what the writer has committed to. Concretely, the inferred layer should present only inferences closely related to the writer's stated intents, with more distant ones held back and offered on demand. The threshold of inference can adapt to the writer: extended more generously when the stated intents are sparse and held back when they are already rich.

\begin{table}[t]
    \caption{Maintaining intent coherence in ordinary co-writing and with \system{}: which steps each process performs, and how the writer's work changes.}
    \label{tab:division}
    \Description{A five-column table. For each of four steps in maintaining intent coherence it marks whether the step is performed in the baseline and in IntentWeaver, then names what the writer's work with IntentWeaver adds and what it lightens. Externalizing and organizing intents happens only in IntentWeaver and adds checking the derived hierarchy. Evaluating the text against intents happens in the baseline only for low-level intents; with IntentWeaver it adds checking the fulfillment labels and lightens judging fit sentence by sentence. Exploring and deciding on changes happens in both; with IntentWeaver it adds confirming each change and lightens exploring directions unaided. Propagating changes and resolving side effects happens only in IntentWeaver and adds checking the impact and conflicts.}
    \centering
    \includegraphics{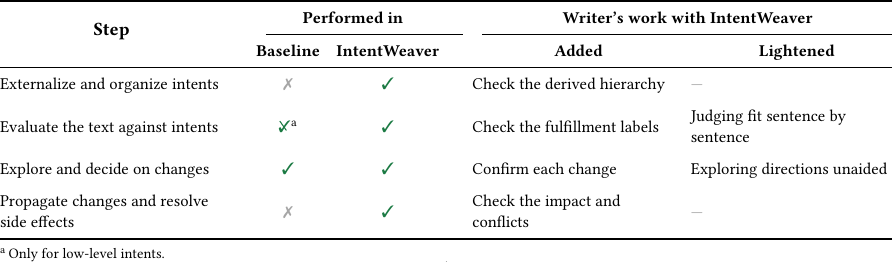}
    \end{table}

\subsubsection{Let the system maintain intent while the writer oversees it}
Table~\ref{tab:division} contrasts the two processes, showing that ordinary co-writing lacks three steps compared with \system{} (organizing intents, evaluating higher-level ones, and propagating each change) that are necessary to keep intent coherent as it evolves. They are missing because a writer absorbed in the draft has no bandwidth left to run them, and asking the writer to perform them, say by prompting them to restate and check their intents each round, would fail on the same grounds. \system{} instead has the system perform all three and asks the writer only for lightweight oversight and confirmation. 
The results indicate that this division of labor worked. 
The three steps took effect (RQ1) while task workload and sense of agency stayed level with the baseline (RQ3), 
which suggests that writers were willing to delegate these steps \cite{lee2004trust,risko2016cognitive}. We therefore offer this division, with the system maintaining intent and the writer overseeing it, as a default for systems of this kind. One qualification concerns evaluation, the step some writers were reluctant to hand over: four participants distrusted the fulfillment labels and wanted to make that judgment themselves. Overseeing this step should therefore include the ability to overrule it: an assessment presented as a contestable claim with its evidence in view, which the writer can revise directly.

\subsubsection{Counter fixation at both levels of intent and text}
A structure derived from the output can only show what the output already does, and so it
binds the writer to the current version. This is the condition the creativity literature calls fixation, a premature convergence on one version that is then refined instead of
reconsidered~\cite{jansson1991design}, and generative AI has been shown to deepen
it~\cite{wadinambiarachchi2024effects}. 
\system{} addresses it at two levels. At the level of text, each draft is regenerated from the writer's committed intents instead of patched from the previous version, so by design no draft is carried forward; this countermeasure, however, requires the writer to re-read the whole draft each round, which adds to their burden (P09, P24). At the level of intent, hints propose intents outside the current draft but tied to those the writer has committed to. This countermeasure is light:
a hint costs nothing to ignore, and ten participants who never adopted one still rated hints
highly, valuing a dimension they \textit{``have to consider''} (P07). 
We therefore recommend that systems grounding intent in a product counter fixation at both levels.    The intent-level countermeasure, a channel of candidate intents that lets users step outside the current version while staying anchored to their own intents (Section~\ref{sec:7.2.1}), is cheap and applies regardless of the product. The text-level countermeasure should scale with the cost of going through the product. Where that cost is low, as with a short text, the product can be remade at every iteration; where it is high, as with a long document or a large codebase, regeneration can be scoped to the parts the changed intents touch, so that the user is not asked to re-read what did not change.

\subsection{Generalizability}
\label{sec:disc-general}

Although developed for co-writing, the interaction logic of \system{} can extend to other
human-AI co-creation tasks, since intent in general is emergent~\cite{stamp1990construct}
and hierarchical~\cite{miller2017plans,carver2001self}, so that the same incoherence arises in many of
them. In agentic coding, accepted local fixes drift from the architecture the specification
set out~\cite{chen2026code}; in data analysis, steps analyzing interesting intermediate results drift from the question originally asked~\cite{kazemitabaar2024improving}; in design, accepted generated details drift from the original brief~\cite{wadinambiarachchi2024effects}. 
The approach suits less well where there is no room for such drift: single-turn requests (e.g., translating), tasks with a goal fixed in advance (e.g., proofreading), and settings where speed matters more than grounding (e.g., replying in live chats).

Also, the mechanisms of \system{} tolerate imperfect model performance, which eases transfer to domains where intent inference is weaker. A derived intent or hint that misses the mark can be left unadopted and affects neither the text nor the writer's other intents (P19, P24). An imprecise warning still made P02 \textit{``think one layer further.''} Many features work by their presence alone, prompting the writer to consider a dimension whether they take it up (P07, P19). What carries over to other domains is therefore not the accuracy of any inference but the arrangement around it.

\subsection{Limitations and Future Work}
\label{sec:disc-limitations}

Our work has several limitations. \textbf{First}, two measures warrant caution. \textit{Drift detection} was measured from participants' judgments during a session replay, and such retrospective reports are open to rationalization after the fact~\cite{nisbett1977telling,ericsson2017protocol}; a complementary measure would code drift directly from the version history against the declared intents. \textit{Ease of use} was measured on first contact with a novel interface, which conflates the cost of learning \system{} with the cost of using it; repeated sessions would let the two be separated. 
\textbf{Second}, our participants were aged 19 to 32. Writers of this generation are accustomed to AI tools and may delegate more readily than writers with settled planning habits. 
\textbf{Third}, the writing tasks were relatively short posts on everyday topics. Participants expected \system{} to offer more for longer, demanding pieces such as academic papers and technical reports (P18--P20, P22--P24), whose intents \textit{``relate to each other in tangled ways that are hard to sort out''} (P24). Future work should examine the approach there, and in other co-creation domains (e.g., agentic coding). 
\textbf{Fourth}, each participant used \system{} for a single 25-minute session. Two effects can only appear under extended use: whether writers can keep track of a large, long-evolving intent structure over many iterations, and whether the knowledge-transforming engagement described in Section~\ref{sec:disc-emergent} yields a lasting learning gain. A multi-week deployment with writers working on their own documents would allow both to be observed.
\textbf{Fifth}, IntentWeaver's graph action space (pin, unpin, edit, and add) excludes setting contribution types and fulfillment levels, to keep every value traceable to text evidence. Future work should explore a larger action space over these attributes and structure, and test whether the added expressiveness repays the added burden.

\section{Conclusion}

In this paper, we present \system{}, an AI co-writing system that treats writing intent as a structure emerging from the text rather than a specification fixed before it. Informed by a formative study, \system{} derives an intent graph from the evolving text, surfaces coherence breakdowns, and propagates the writer's changes across the hierarchy. A within-subjects study showed participants returned to their high-level intents more often, evolved them further, and produced final texts that better realized those intents. We further identified distinct patterns of engaging with system and distilled design considerations for treating intent as a live structure. More broadly, this work demonstrates that the product of human-AI collaboration can itself become the medium through which intent is discovered.

\bibliographystyle{ACM-Reference-Format}
\bibliography{CHI27}

\end{document}